\documentclass{elsart}

\usepackage[square]{natbib}

\usepackage[scriptsize]{subfigure}
\usepackage{graphicx}
\graphicspath{{figures/}}

\usepackage{amssymb}
\journal{Nucl. Instr. and Meth. in Phys. Res. A}

\def\KAOS{{\sc Kaos}$/\!${\small A1}\@}
\def\PANDA{$\overline{\mbox{P}}${ANDA}\@}
\renewcommand{\ell}{l}

\begin{document}
\begin{frontmatter}

\title{Characterisation of radiation damage in silicon photomultipliers
	with a Monte Carlo model}

\author{S.~{S\'anchez Majos}\thanksref{PhD}\corauthref{cor}},
\ead{sanchez@kph.uni-mainz.de} \author{P.~Achenbach}, and
\author{J.~Pochodzalla}

\address{Institut f\"ur Kernphysik, Johannes Gutenberg-Universit\"at
    Mainz, Germany}

\thanks[PhD]{Part of doctoral thesis.}
\corauth[cor]{Corresponding author. Tel.: +49-6131-3925851;
  fax: +49-6131-3922964.}

\begin{abstract}
  Measured response functions and low photon yield spectra of silicon
  photomultipliers (SiPM) were compared to multi-photoelectron
  pulse-height distributions generated by a Monte Carlo
  model. Characteristic parameters for SiPM were derived. The devices
  were irradiated with 14\,MeV electrons at the Mainz microtron
  MAMI. It is shown that the first noticeable damage consists of an
  increase in the rate of dark pulses and the loss of uniformity in
  the pixel gains.  Higher radiation doses reduced also the photon
  detection efficiency. The results are especially relevant for
  applications of SiPM in fibre detectors at high luminosity
  experiments.
\end{abstract}


\end{frontmatter}

\section{Introduction}
After a few years of R\&D, silicon photomultipliers (SiPM) are today
an interesting alternative to conventional vacuum phototubes in many
applications due to the specific advantages of a solid state device:
small size, low voltage operation and magnetic field
insensitivity~\cite{Buzhan2001,Dolgoshein,Dolgoshein-2006}. Nevertheless,
the idea of grouping hundreds of miniature avalanche photodiodes (APD)
in a planar array to form an analog device has found a limitation in
the high rate of intrinsic dark pulses of such a configuration. Only
small active area devices are currently available and typical noise
rates are of the order MHz$/$mm$^2$ with signal pile-up due to optical
cross-talk. This leads to the situation of a small photo-sensor that
needs to detect a relatively large amount of light in order to achieve
an acceptable signal-to-noise ratio. To our knowledge there has been
no attempt to use SiPM in applications where only a few photons must
be detected. This is for example the situation faced when using
scintillating fibres as tracking detectors.

There is little information about radiation hardness of SiPM. It is
well known from other types of silicon detectors that generating
centres are created during irradiation which increase the leakage
current~\cite{Leroy}. The bulk leakage current is multiplied in APD by the
gain factor and the resulting pulses are undistinguishable from photon
generated events. Consequently, an increasing rate of dark
pulses as a function of the radiation dose is expected for SiPM. Low
light level detection will be degraded or even impossible in some
applications if this effect happens to be large. Adverse effects of
irradiation on other characteristic parameters of SiPM such as gain
uniformity, after-pulsing or optical cross-talk probability will be
also detrimental for a detector. In general, it is mandatory to study
the impact of the particular kind of radiation the detector will be
exposed to on the characteristic parameters that must remain stable.

In this paper a complete model for multi-photoelectron pulse-height
distributions (MPHD) was used to extract the main consequences of
electromagnetic irradiation on SiPM. Section~2 discusses the MPHD
model. Section~3 gives a brief description of the irradiation
set-up. Section~4 describes the changes in SiPM characteristics
obtained for low and high irradiation doses.  Section~5 closes with
future prospects of using this kind of detectors in low light level
applications.

\section{The Monte Carlo model for the detector output}
A complete model for the MPHD of SiPM was derived based on a Monte
Carlo simulation. For the model, Poisson distributed photo-electron
and dark signals are generated independently. Each of these signals
can cause optical cross-talk and after-pulses according to a given
probability distribution. For this set of pixels the distribution in
time is generated with respect to an integration time window. Single
pixels can contribute with varying gain to the generation of the
detector output. Finally, the noise is added. The free parameters of
the simulation are: mean signal amplitude ($A$), gain variation
($\sigma_{G}$), dark pulse rate ($r$), optical cross-talk probability
($p_{\it opt}$), after-pulse probability ($p_{\it aft}$), trap
lifetime ($\tau$), mean number of detected photons ($\lambda$),
pedestal position ($x_{ped}$), and noise amplitude
($\sigma_{ped}$). In the following, the simulated processes are
explained.

As SiPM are a set of APD connected in parallel many of their
properties are thus inherited. Its dark count rate, $r$, is the sum of
all the APD dark count rates.  After-pulses appear in each pixel after
a photon (or a thermically generated charge carrier) triggers an
avalanche due to trapped carriers that are released after some
time-delay. The after-pulse probability, $p_{aft}$, is defined as the
fraction of events in which one additional signal is generated after
the detection of a photon. It is the product of the trapping
probability and the triggering probability, both increasing linearly
with the applied bias voltage. The trap lifetime, $\tau$, determines
the typical time-scale.
  
SiPM are manufactured so that signal uniformity from pixel to pixel is
quite good, typically within 10\,\%~\cite{Renker}. The small
variation, $\sigma_{G}$, together with the narrow single electron
response function of each APD provides excellent photon counting
capabilities and as many as 20 photons can easily be resolved in a
typical pulse height spectrum~\cite{Buzhan2001}. The distance between
multi-pixel peaks in the spectrum is a measure of the charge gain,
$A$.

Surface leakage currents do not cross the multiplication region. Their
fluctuations are merged with other noise sources to define the noise
amplitude as measured by the pedestal width, $\sigma_{ped}$. A change
in the surface leakage current should also appear as a shift in
pedestal position, $x_{ped}$.

\begin{figure}
  \centering
  \includegraphics[width=0.6\textwidth]{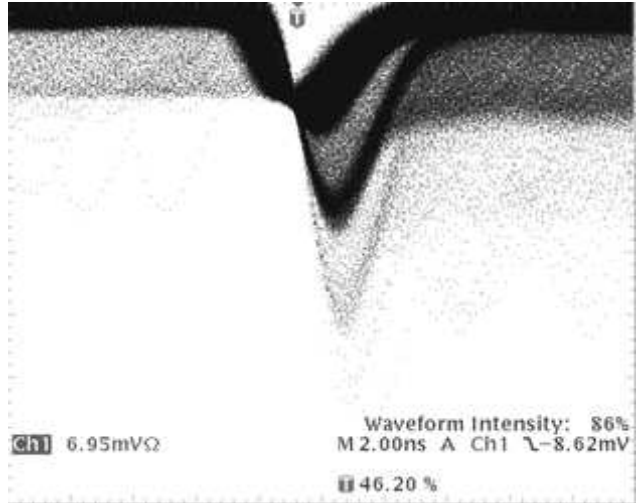}
  \caption{The simultaneous signals from up to three pixels are
    visible in this oscilloscope trace. Optical cross-talk between
    adjacent pixels that form a cluster generates these events. The
    MPHD model described in the text assumes all signals in a cluster
    to appear at the same time.}
  \label{fig:optcrosstalk}
\end{figure}

An avalanche of $10^6$ carriers in any of the micro-metric APD forming
the SiPM will create around 50 photons via hot carrier luminescence
with enough energy to trigger any neighbouring
pixel~\cite{Otte}. Devices without trenches filled with opaque
material exhibit optical cross-talk, where at least one photon is able
to cross the spacing between micro-cells to produce a simultaneous
signal (within 100\,ps). Fig.~\ref{fig:optcrosstalk} shows the effect
of optical cross-talk in the SiPM elementary signal. The well defined
amplitudes of the multi-pixel events is a result of the simultaneity
of the composing signals.

To the knowledge of the authors there has been only one serious
attempt to describe the statistics of multi-photoelectron pulse-height
distributions~\cite{Wright}. A.G.~Wright points out that the
combination of the Poissonian nature of most light sources and the
Binomial character of the photoelectric effect will give rise to a
Poisson distribution for the number of detected photons. He goes on by
establishing a method that uses this fact in an efficient way to build
up the full pulse-height spectrum by using the response function of
the corresponding detector.  This method faces the fact that in many
cases the response to a single photon is far from being
Gaussian. Experimental input is used to accurately describe the
amplitude distribution with no ad hoc assumption about how the
detector will react to a single photon.

\begin{figure}
  \centering 
  \subfigure[Response function]{\includegraphics[width=
    0.48\textwidth]{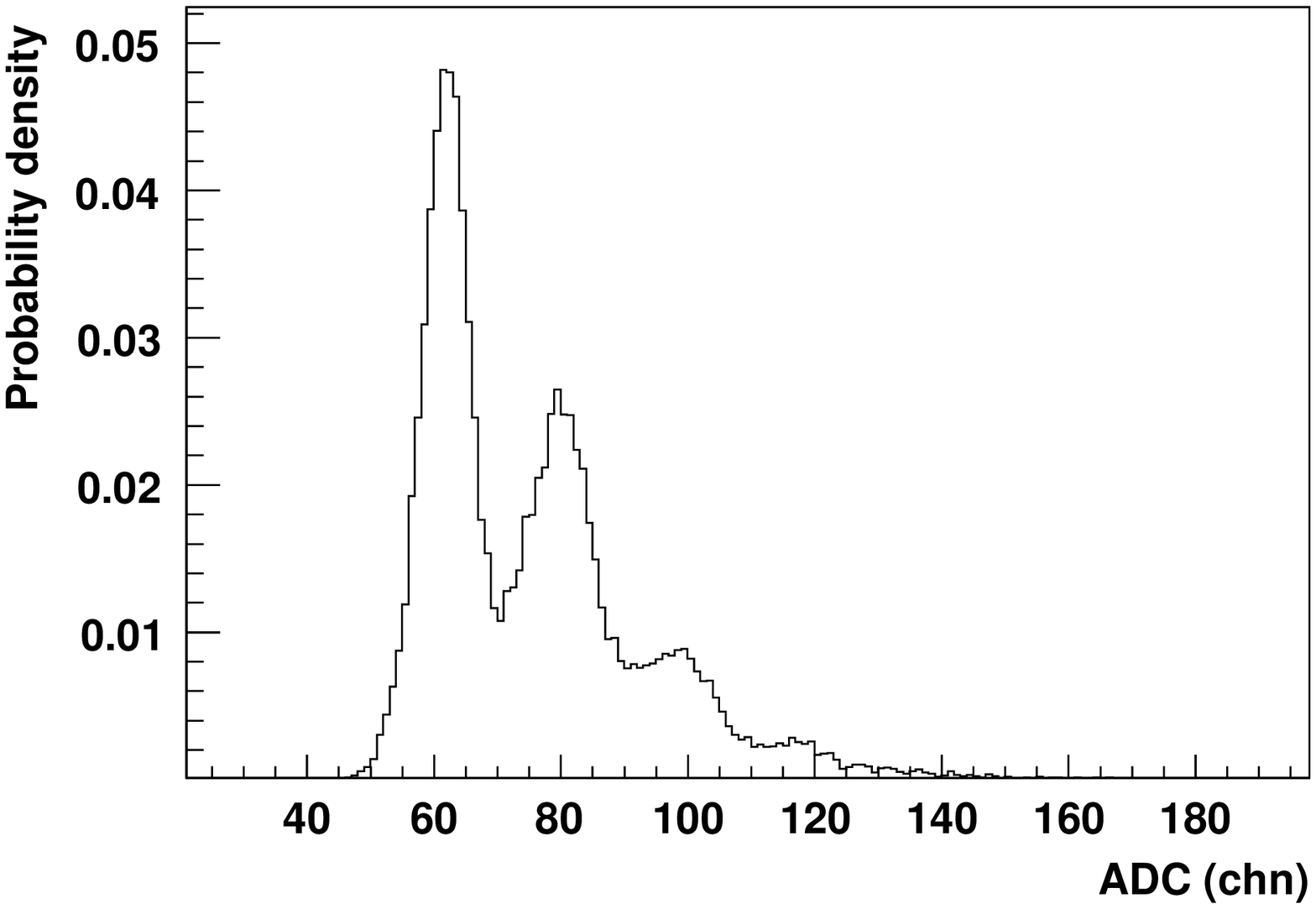}
    \label{fig:resp_func_a}}
  \subfigure[Noise spectrum]{\includegraphics[width=
    0.48\textwidth]{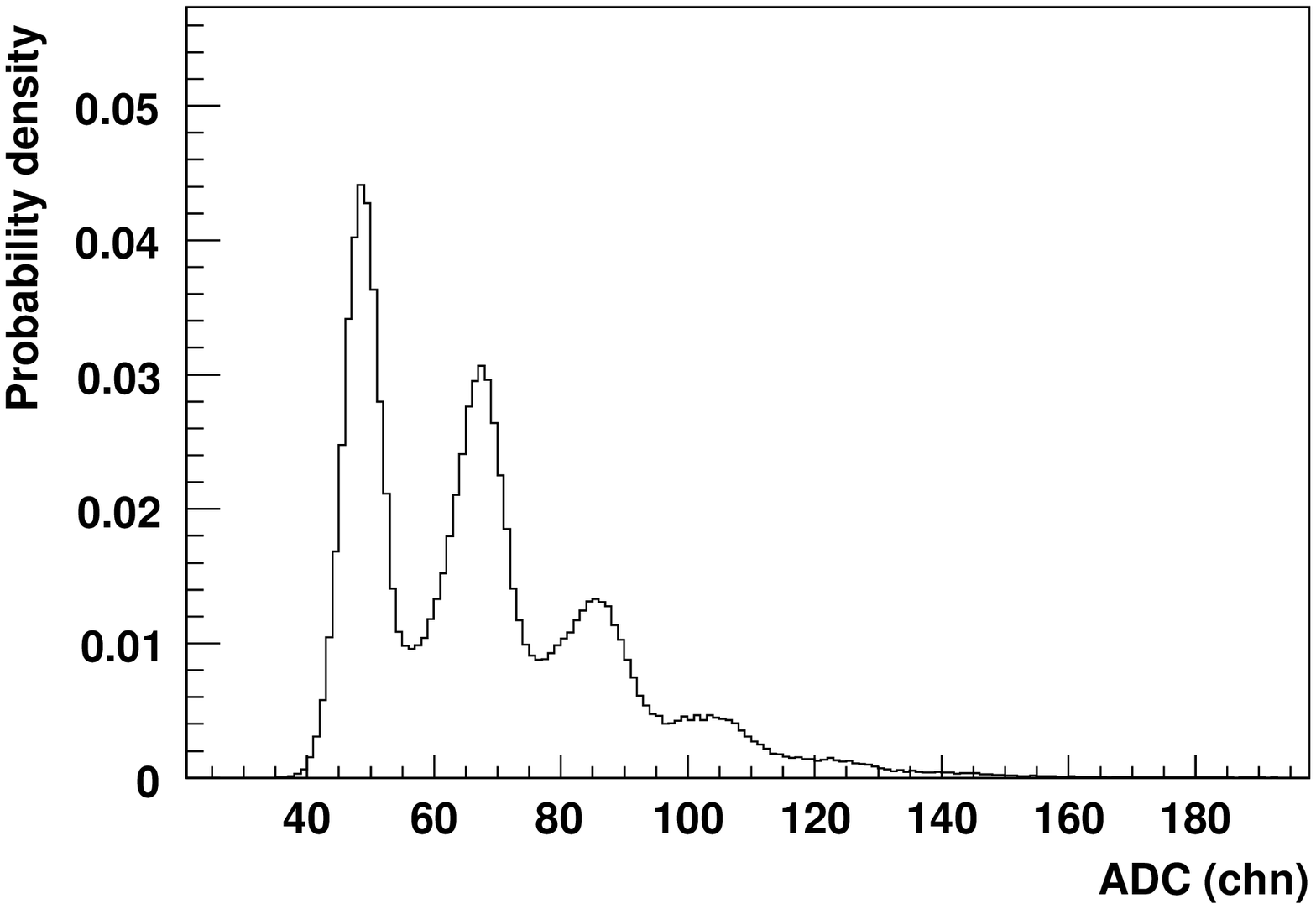}
    \label{fig:resp_func_b}}
  \caption{(a) Measured single-electron response function of a SiPM.
    The multiple peak structure is to a large extend due to the high
    rate of dark pulses.  Optical cross-talk and after-pulses
    contribute to a lesser extend. (b) Integration of noise signals in
    a SiPM with a randomly generated gate. Events outside the pedestal
    peak are a result of the high rate of dark pulses.}
\end{figure}

In our MPHD model the violation of Poissonian statistics arises from
the well known SiPM characteristics, incorporating the multiple-peak
structure.  It also makes use of the simplifying assumption that the
individual peaks in the response function are Gaussian distributed.
It was observed that when bias voltages are low and pixels are assumed
to behave in a practically independent way, the Gaussian hypothesis
reproduces with great accuracy the measured spectra.
Fig.~\ref{fig:resp_func_a} shows the measured response function of a
Photonique device\footnote{SSPM-0701BG-TO18 by {\sf Photonique SA},
  http://www.photonique.ch (2007)} for a bias voltage of 17.5\,V after
irradiation with a fluence of $3.1 \times 10^9$ electrons of 14\,MeV
energy.  Fig.~\ref{fig:resp_func_b} shows the result of a random
integration of noise signals for the same gate width. The relatively
large number of counts outside the pedestal peak is a result of the
high rate, $r$, of dark pulses. A simple convolution would add those
events several times. SiPM are evolving toward the situation in which
this algorithm can be successfully applied. Nowadays, devices are
unfortunately not perfect and the high noise rate, for instance, would
make the method inoperative.

\begin{figure}
  \centering
  \subfigure[$4q^3(1-q)^8$]{
    \includegraphics[width=0.23\textwidth]{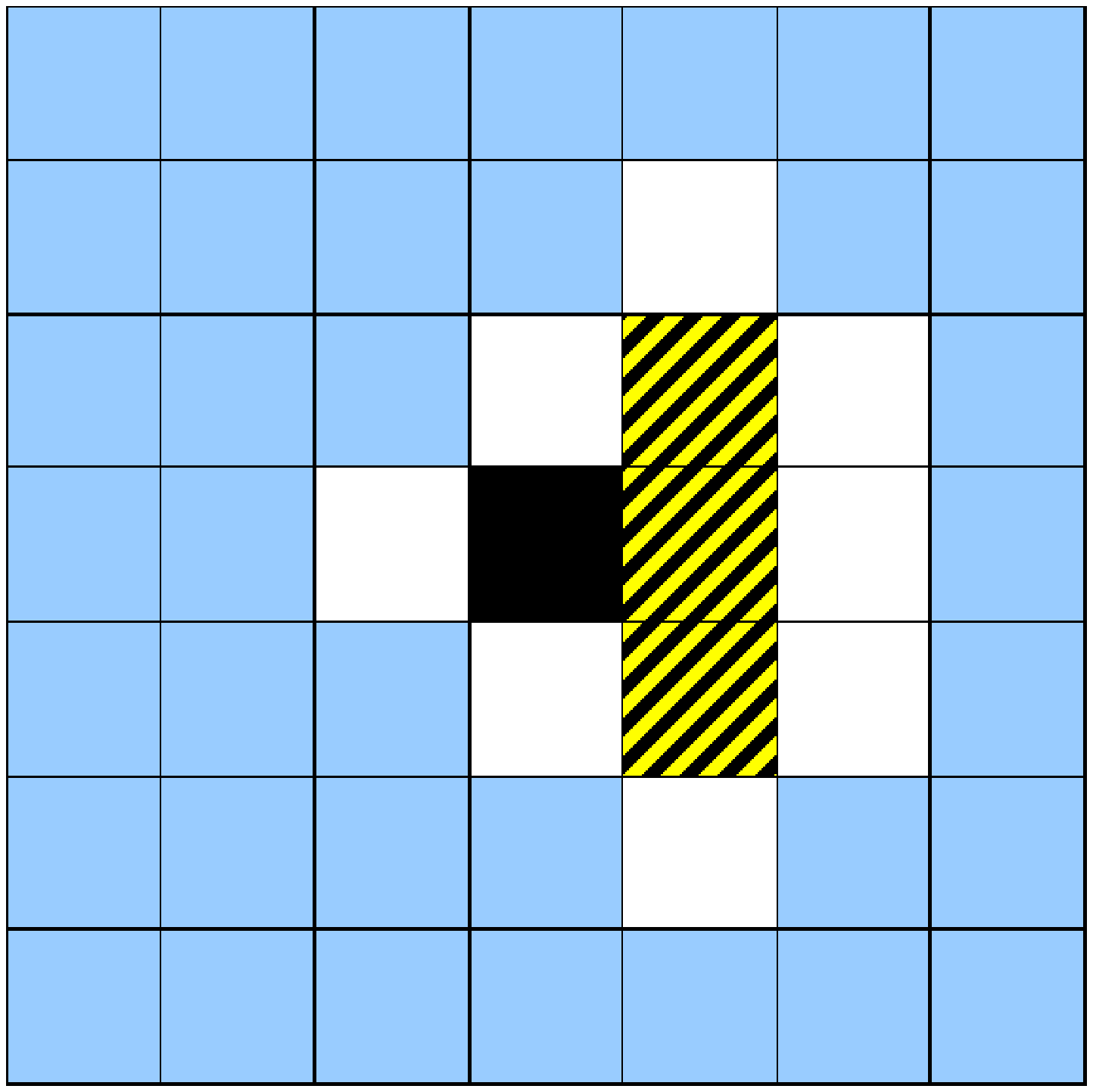}}
  \subfigure[$8q^3(1-q)^9$]{
    \includegraphics[width=0.23\textwidth]{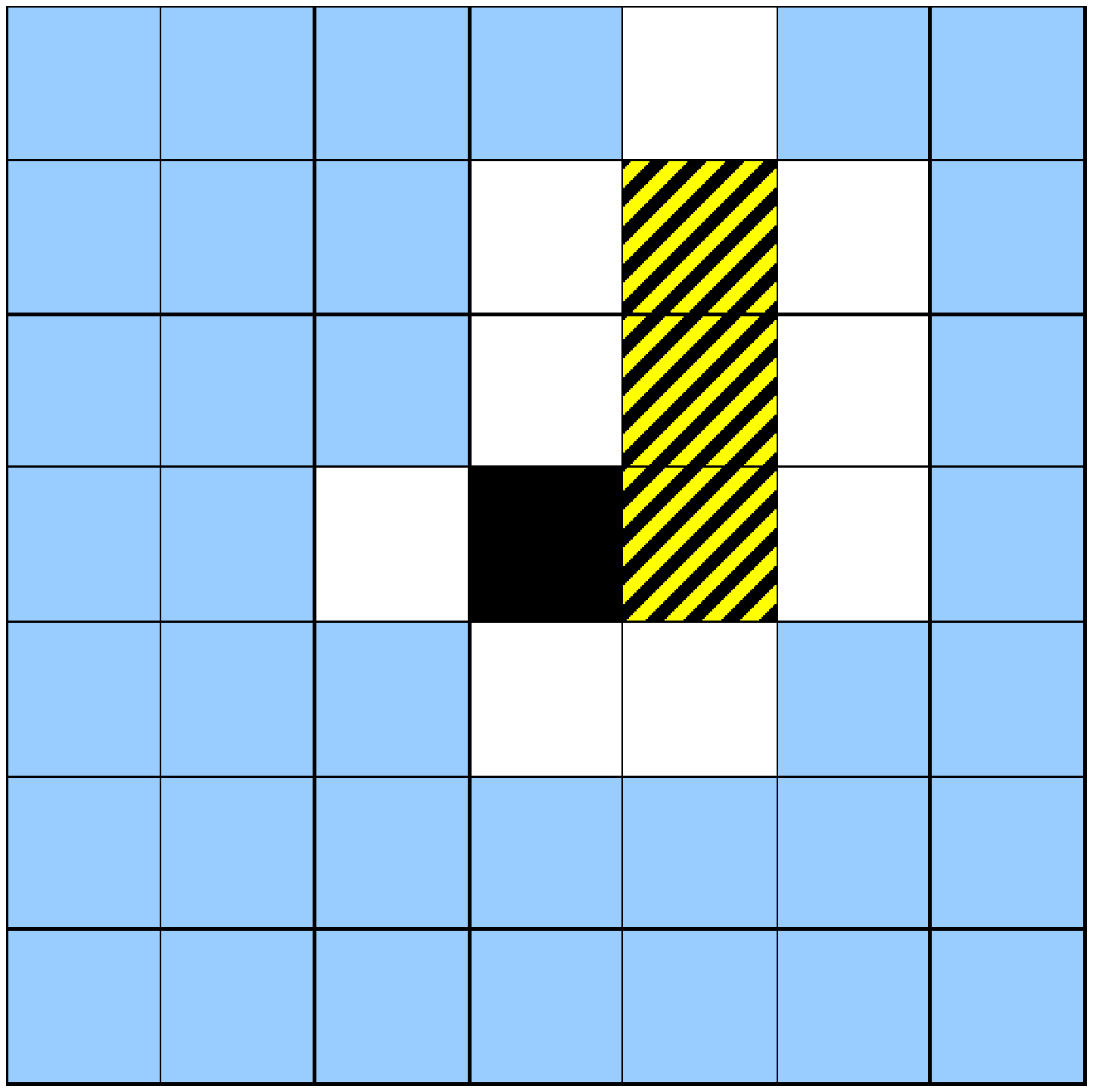}}
  \subfigure[$8q^3(1-q)^9$]{
    \includegraphics[width=0.23\textwidth]{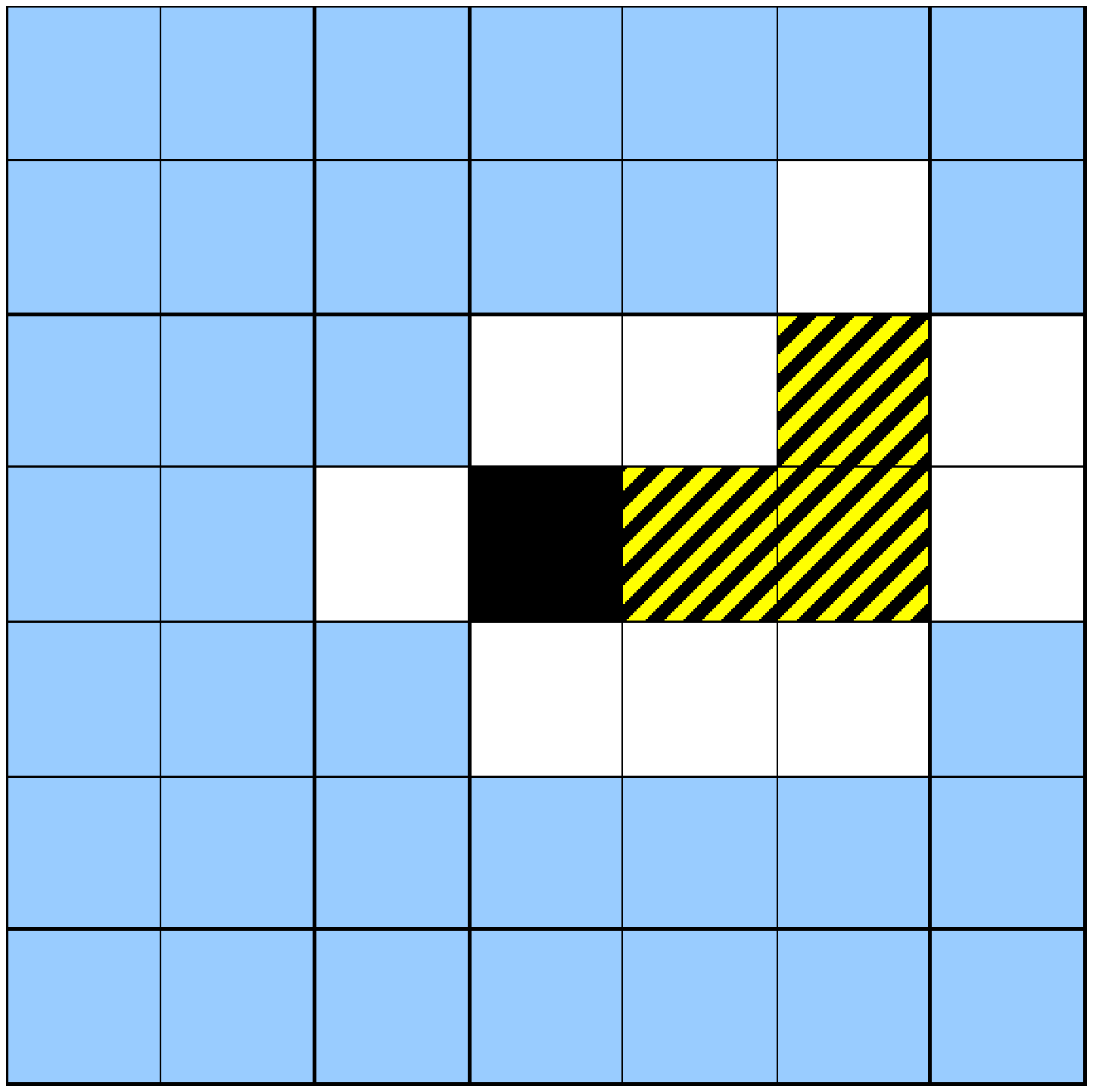}}
  \subfigure[$4q^3(1-q)^{10}$]{
    \includegraphics[width=0.23\textwidth]{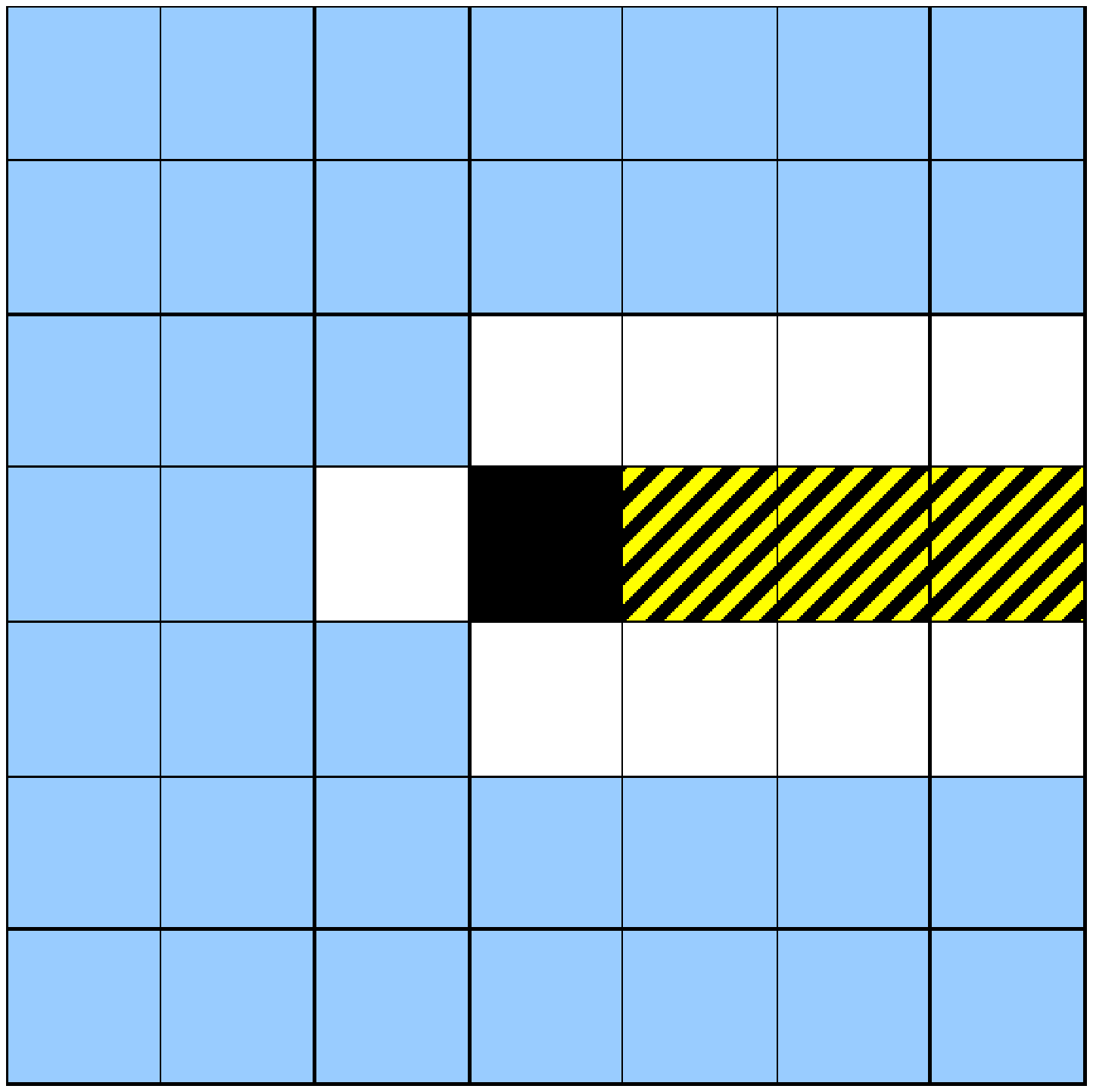}}\\
  \subfigure[$8q^3(1-q)^8$]{
    \includegraphics[width=0.23\textwidth]{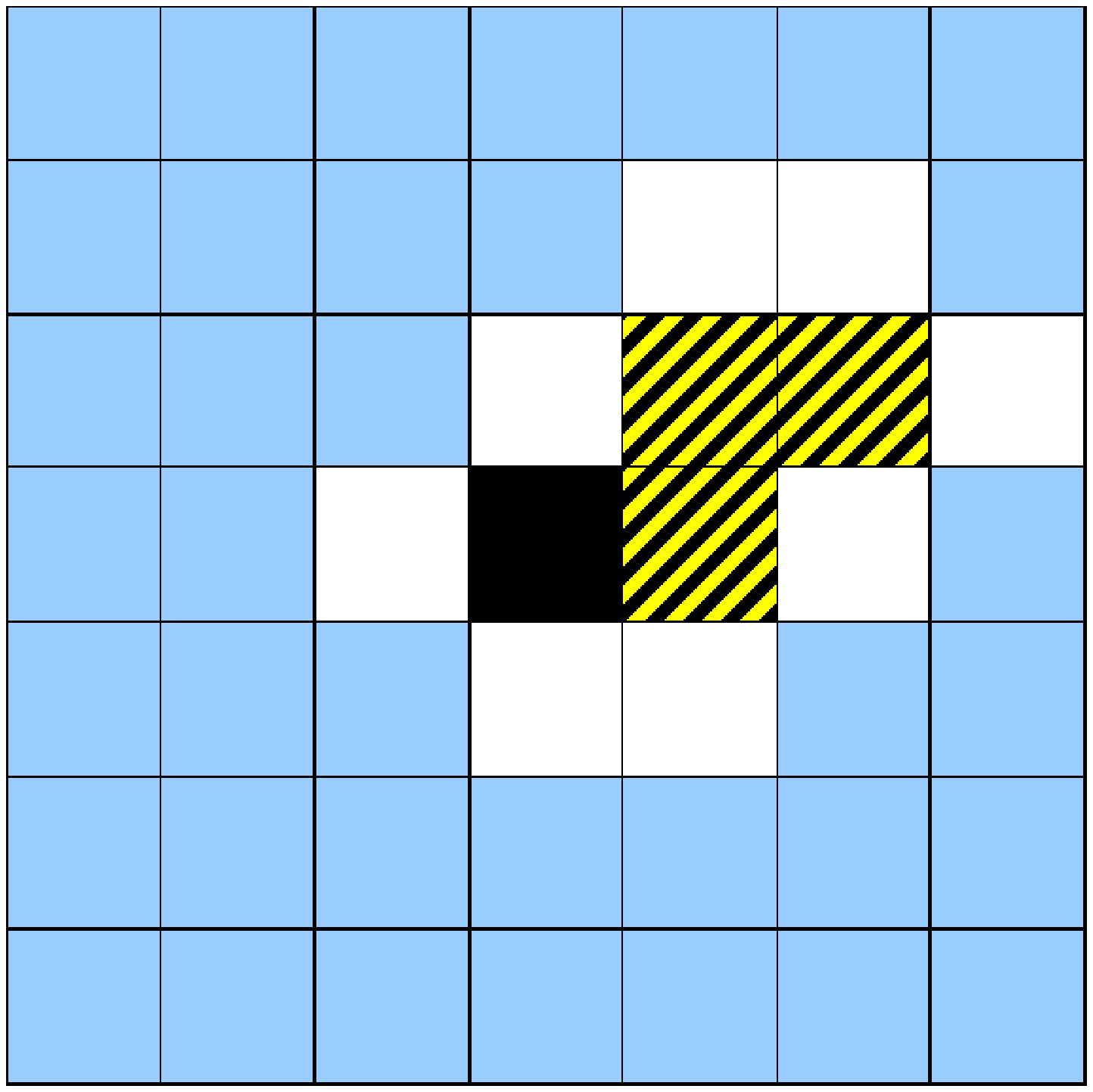}}
  \subfigure[$4q^3(1-q)^{10}$]{
    \includegraphics[width=0.23\textwidth]{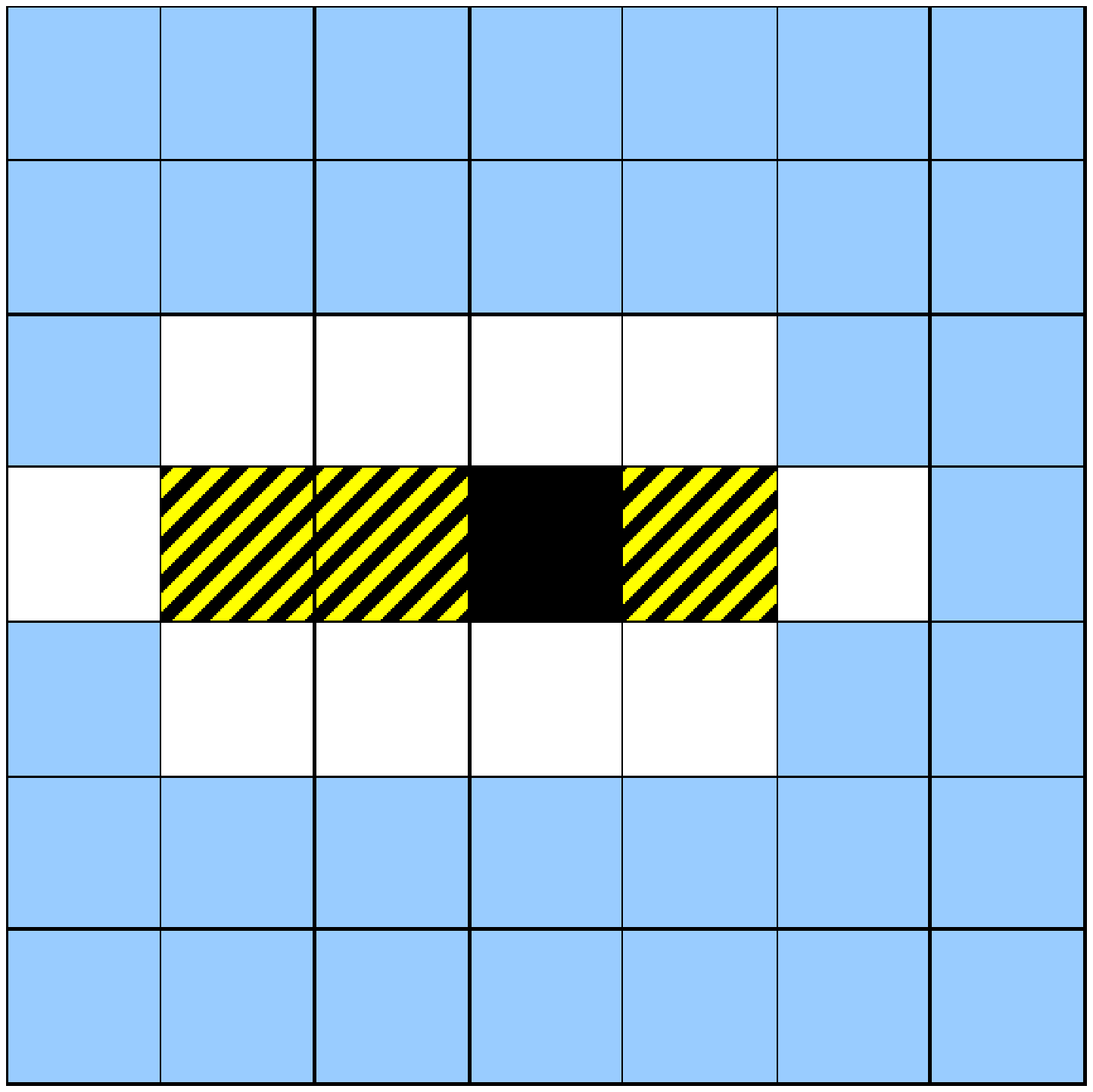}}
  \subfigure[$8q^3(1-q)^9$]{
    \includegraphics[width=0.23\textwidth]{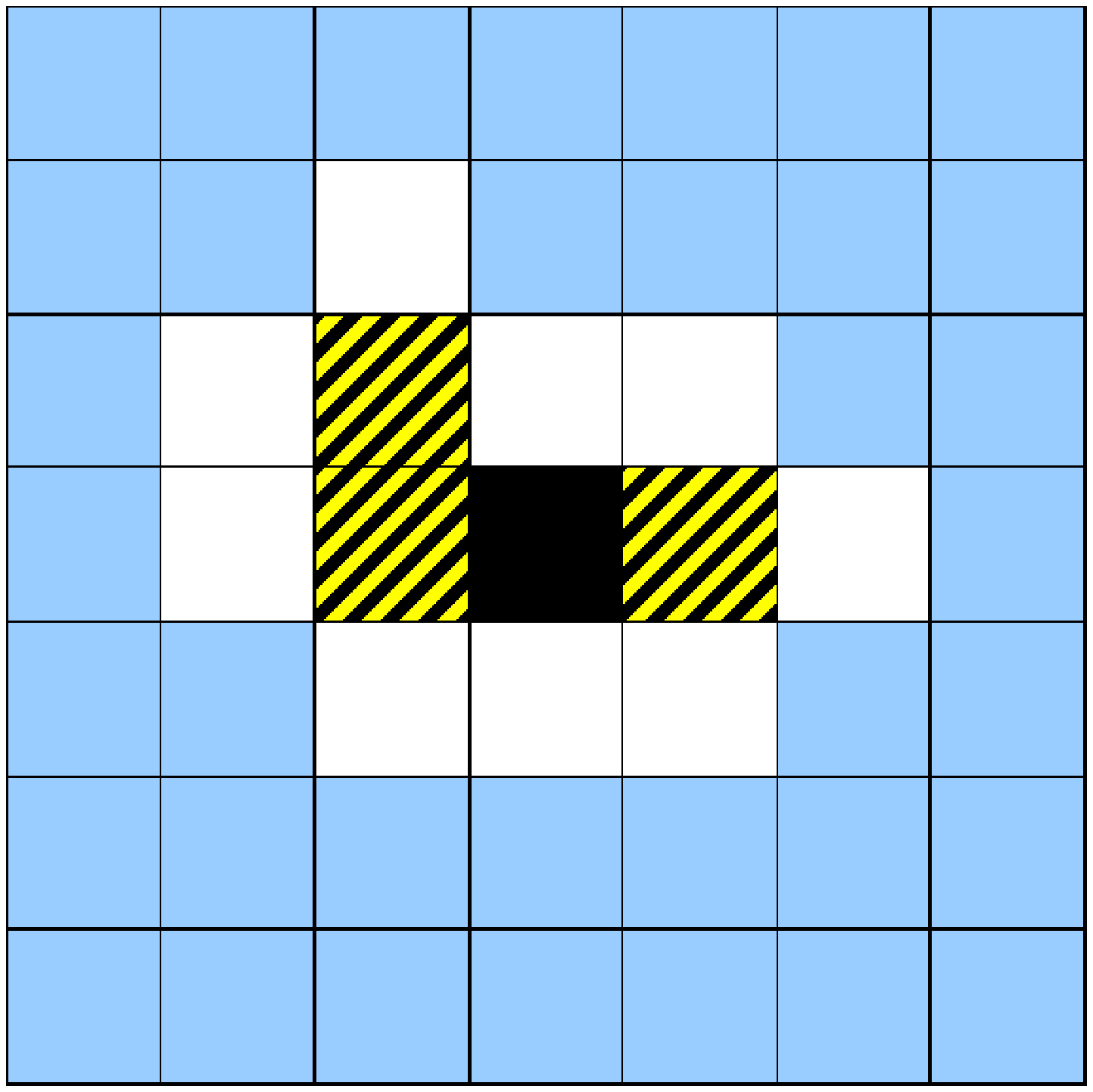}}
  \subfigure[$4q^3(1-q)^8$]{
    \includegraphics[width=0.23\textwidth]{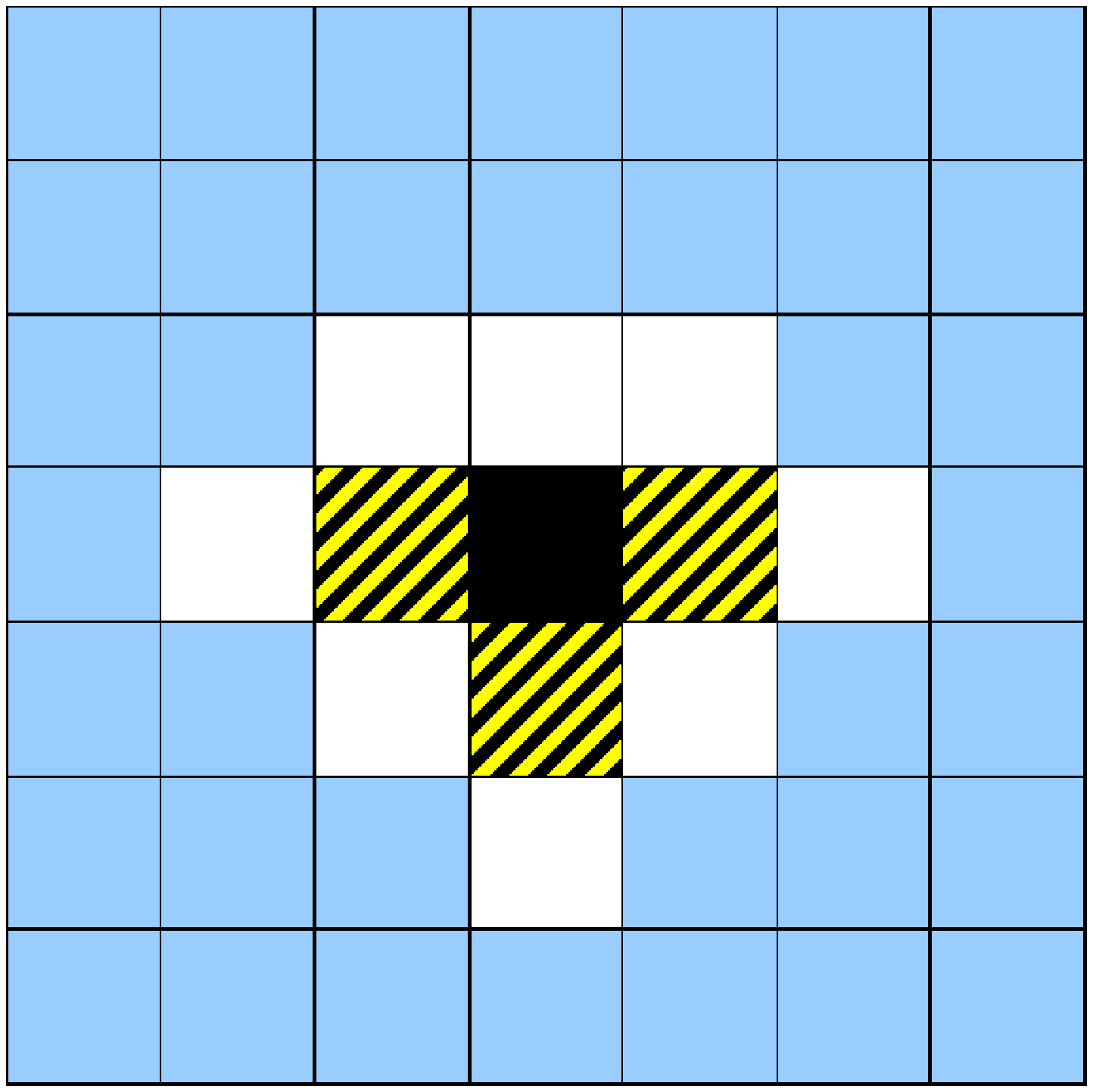}}\\
  \subfigure[$8q^3(1-q)^8$]{
    \includegraphics[width=0.23\textwidth]{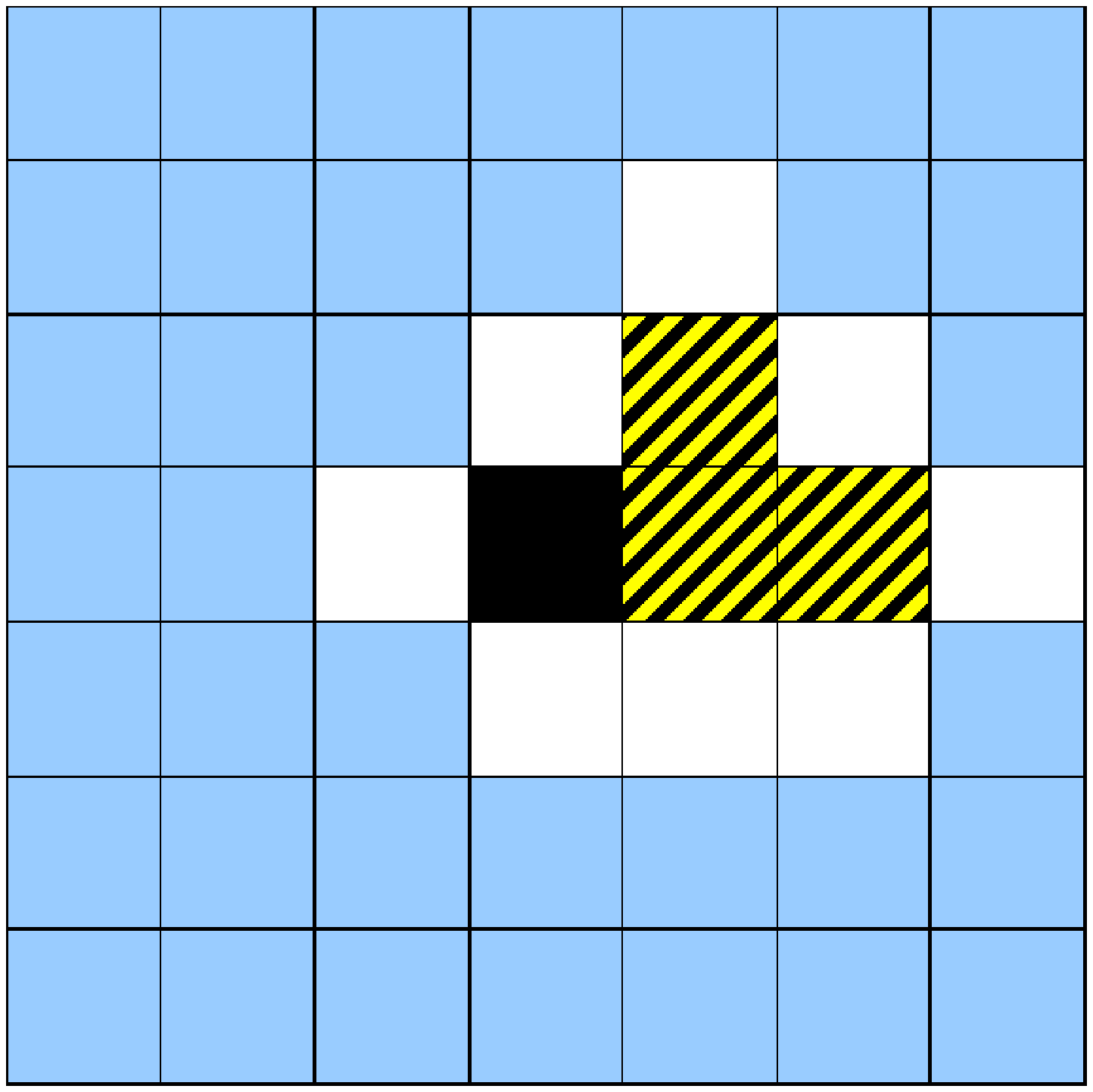}}
  \subfigure[$4q^3(1-q)^8$]{
    \includegraphics[width=0.23\textwidth]{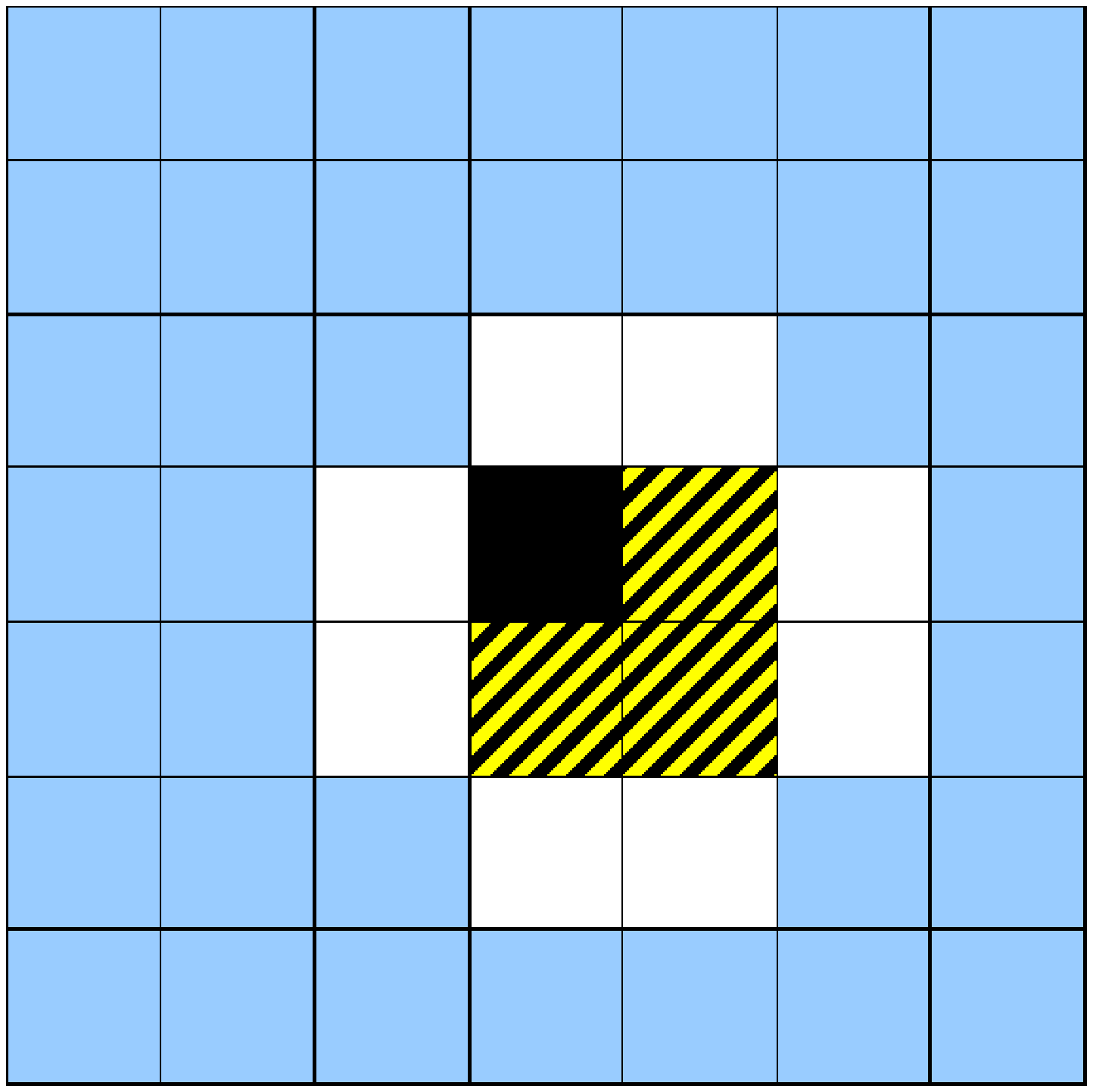}}
  \subfigure[$8q^3(1-q)^8$]{
    \includegraphics[width=0.23\textwidth]{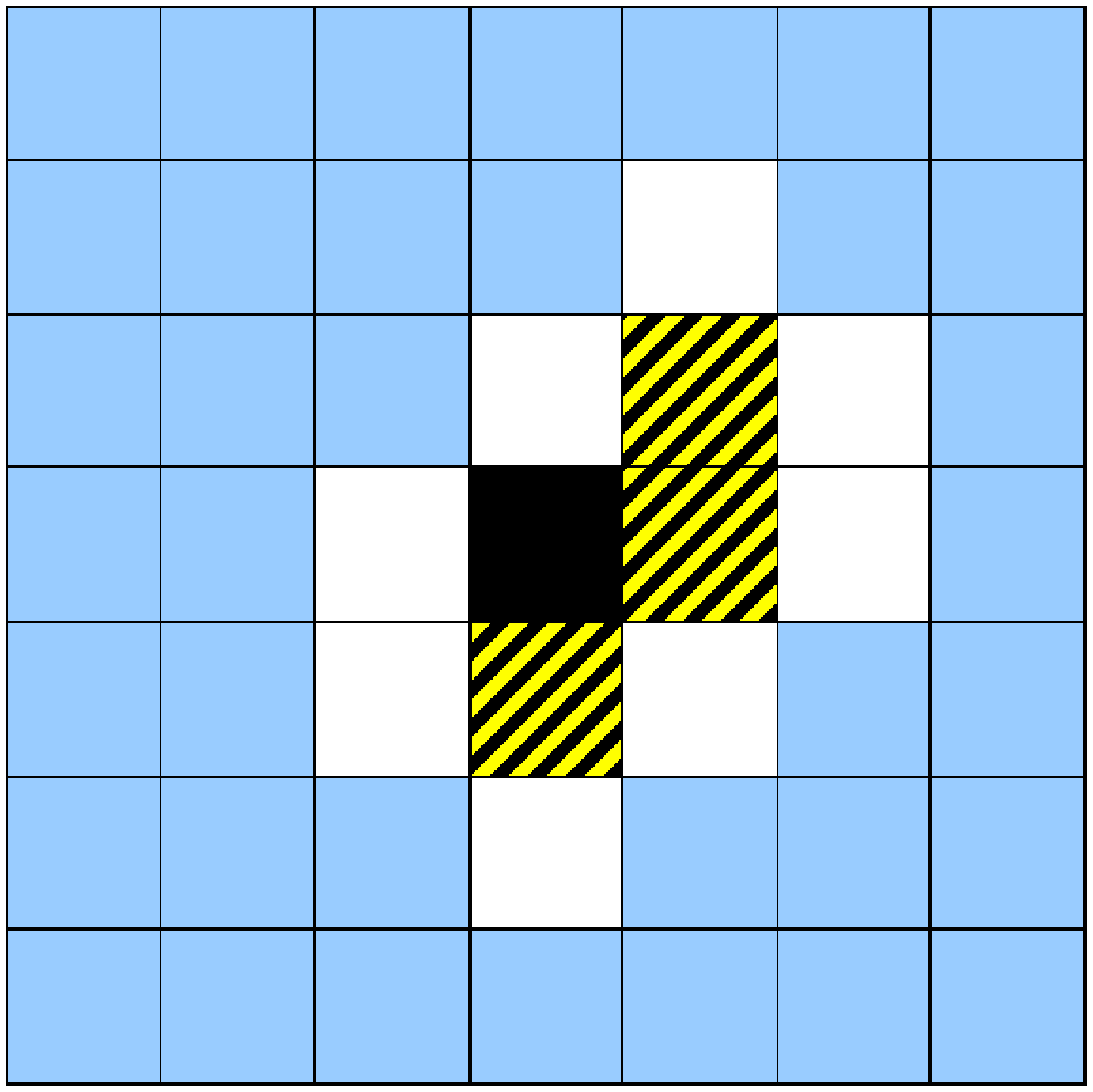}}
  \subfigure[$8q^3(1-q)^9$]{
    \includegraphics[width=0.23\textwidth]{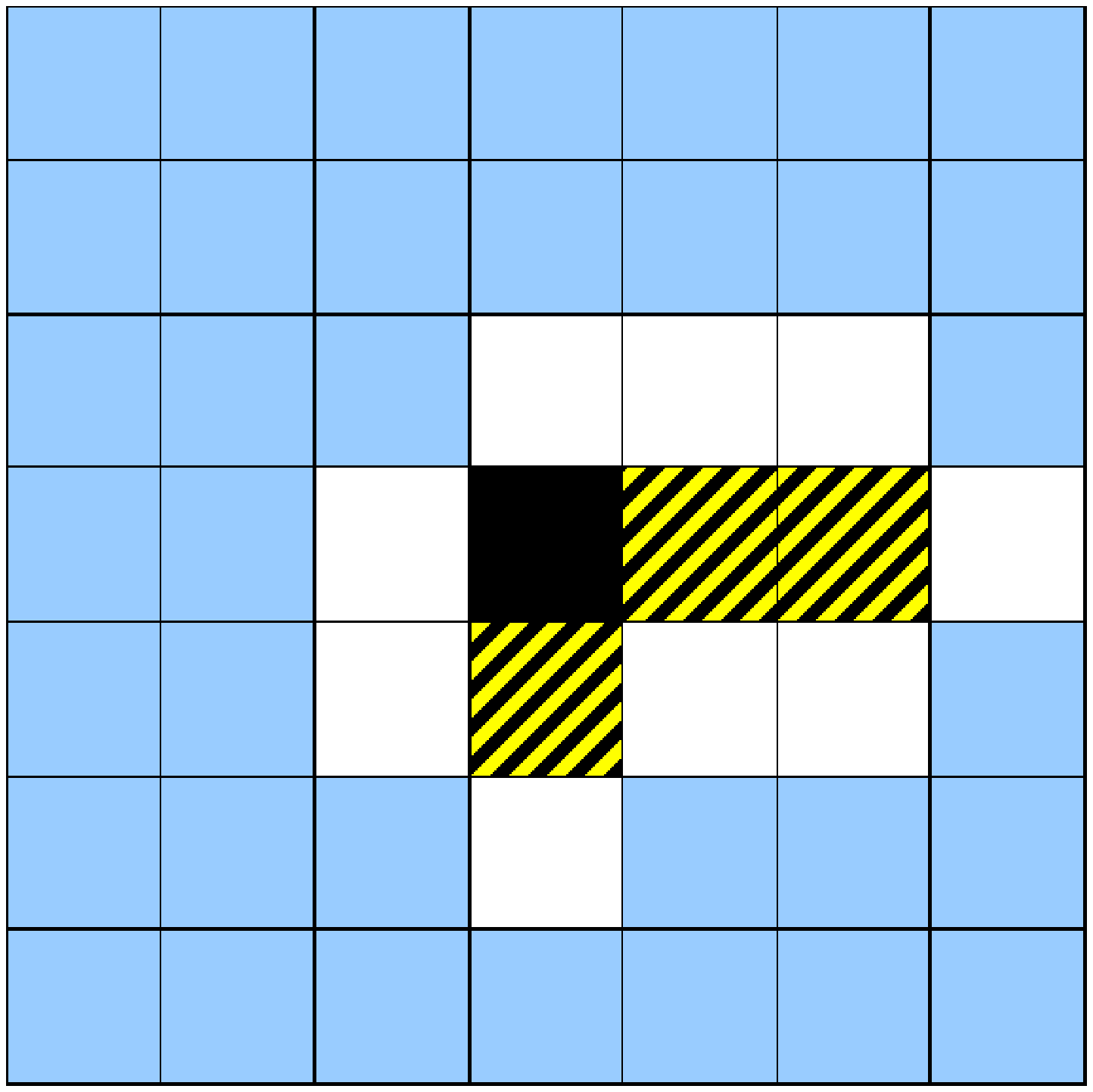}}
  \caption{Independent clusters in which three pixels have been fired
    (crossed cells) in addition to the initial one (solid cell) as a
    consequence of optical cross-talk. Any of the possible 76 patterns
    can be derived from (a) to (l) by symmetry operations. White cells
    represent pixels which remain off. The sum of all terms equals
    $P(3)= q^3[36(1-q)^8+32(1-q)^9+8(1-q)^{10}]$, where $q$ is the
    probability of activation of a single pixel. No distinction is
    made when there is more than one neighbour active.}
  \label{fig:clusters}
\end{figure}

Optical cross-talk is modeled by considering the probability, $q$, of
simultaneous activation of two isolated pixels when one of them has
been triggered and deducing the probability for all possible clusters
in the SiPM pixel matrix. No distinction is made when there is more
than one neighbour active.  Fig.~\ref{fig:clusters} shows all
independent cluster types for the case of three additional
pixels. Only pixels sharing one side are allowed as part of a
cluster. Cluster probabilities are given by the zero pixel cross-talk
probability $P(0)=(1-q)^4$, and the $N$-pixel cross-talk probabilities
$P(1)=4q(1-q)^6$, $P(2)=q^2[6(1-q)^8+12(1-q)^7]$, and
$P(3)=q^3[32(1-q)^8+32(1-q)^9+8(1-q)^{10}]$. The terms with $(1-q)$
are a consequence of the requirement that the pixels outside the
cluster have to remain off.  In absence of optical cross-talk, noise
and after-pulses the pixel number distribution should be
Poissonian. The mean value of this distribution, $\lambda$, gives a
measure of the photon detection efficiency.

\begin{figure}
  \centering
  \includegraphics[width=\textwidth]{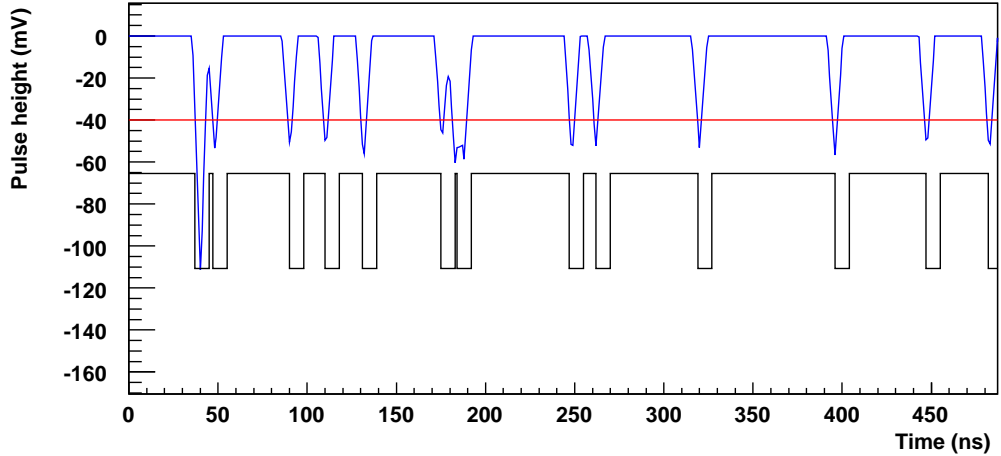}
  \caption{Signals generated by the MPHD
    model. The straight line is the variable threshold level. Output
    of the leading edge discriminator is shown below. Superposition
    of signals and optical cross-talk are visible in the first two
    peaks, a variation in the amplitude is observed for the
    next three single-pixel signals an after-pulse is visible in the 
    next group of piled-up signals.}
  \label{fig:MC_trace}
\end{figure}

A realistic leading edge discriminator was implemented
taking into account the blocking time of the module while the output
is active (10\,ns). Fig.~\ref{fig:MC_trace} shows some of the pulses
generated by the MPHD model, the threshold line and the output of the
leading edge discriminator. Superposition of signals and optical
cross-talk are visible in the first two peaks, a variation in
amplitude can be observed in the next three single pixel signals and
an after-pulse is visible in the next group of piled-up signals.

\begin{figure}
  \subfigure[17.0\,V]{
  	\includegraphics[width=0.48\textwidth]{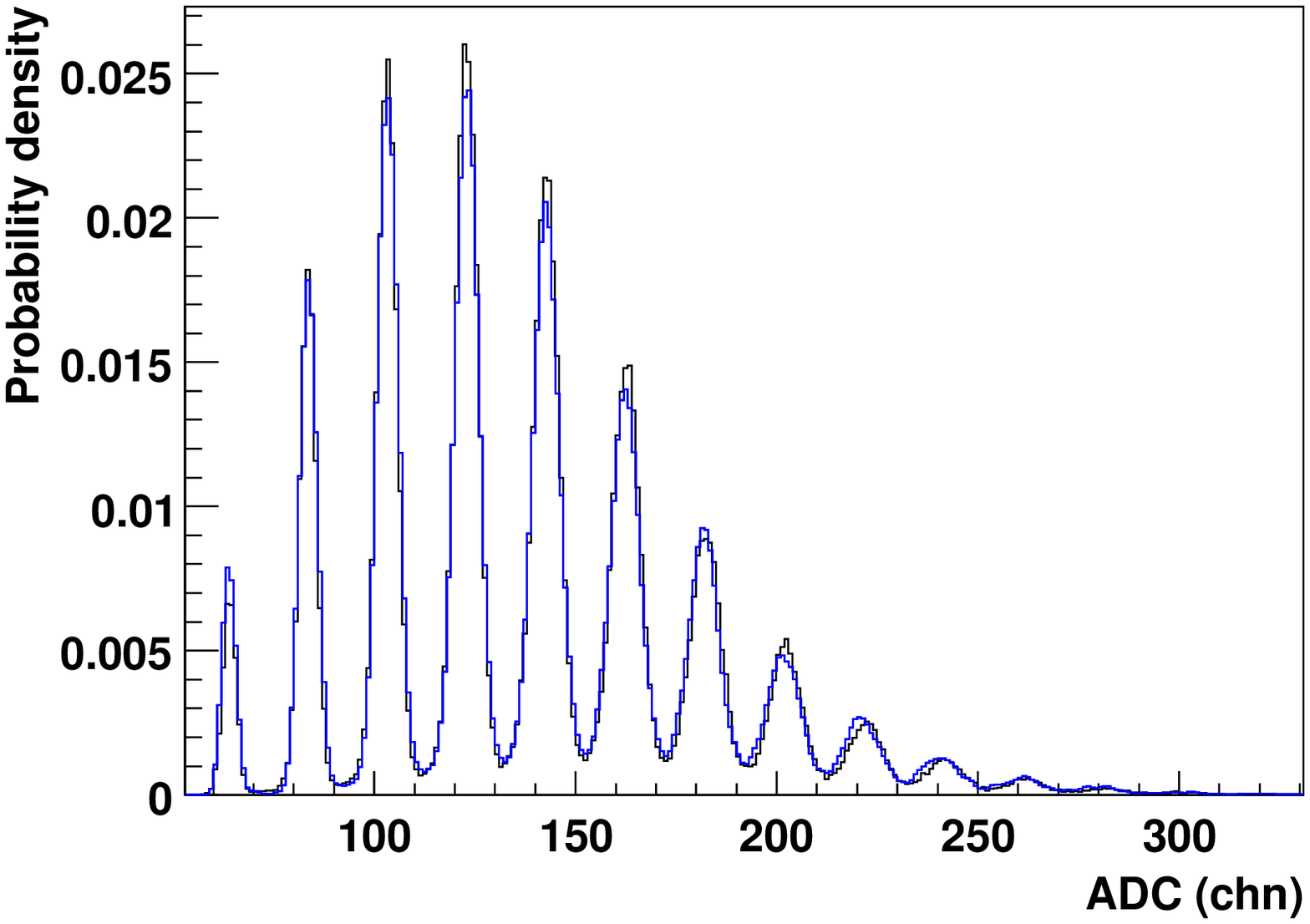}}\hfill
  \subfigure[17.4\,V]{
    \includegraphics[width=0.48\textwidth]{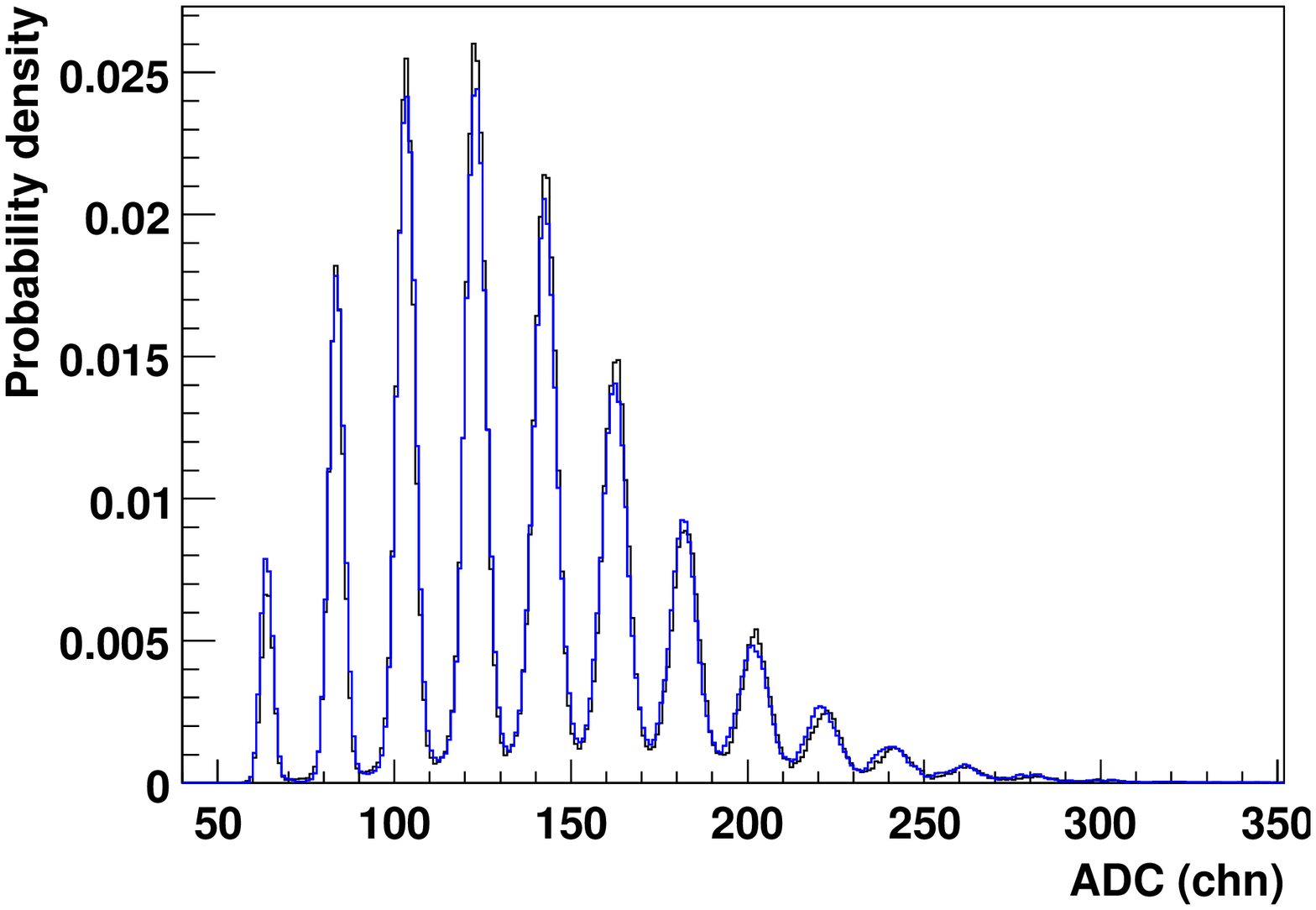}}\\
  \subfigure[17.9\,V]{
    \includegraphics[width=0.48\textwidth]{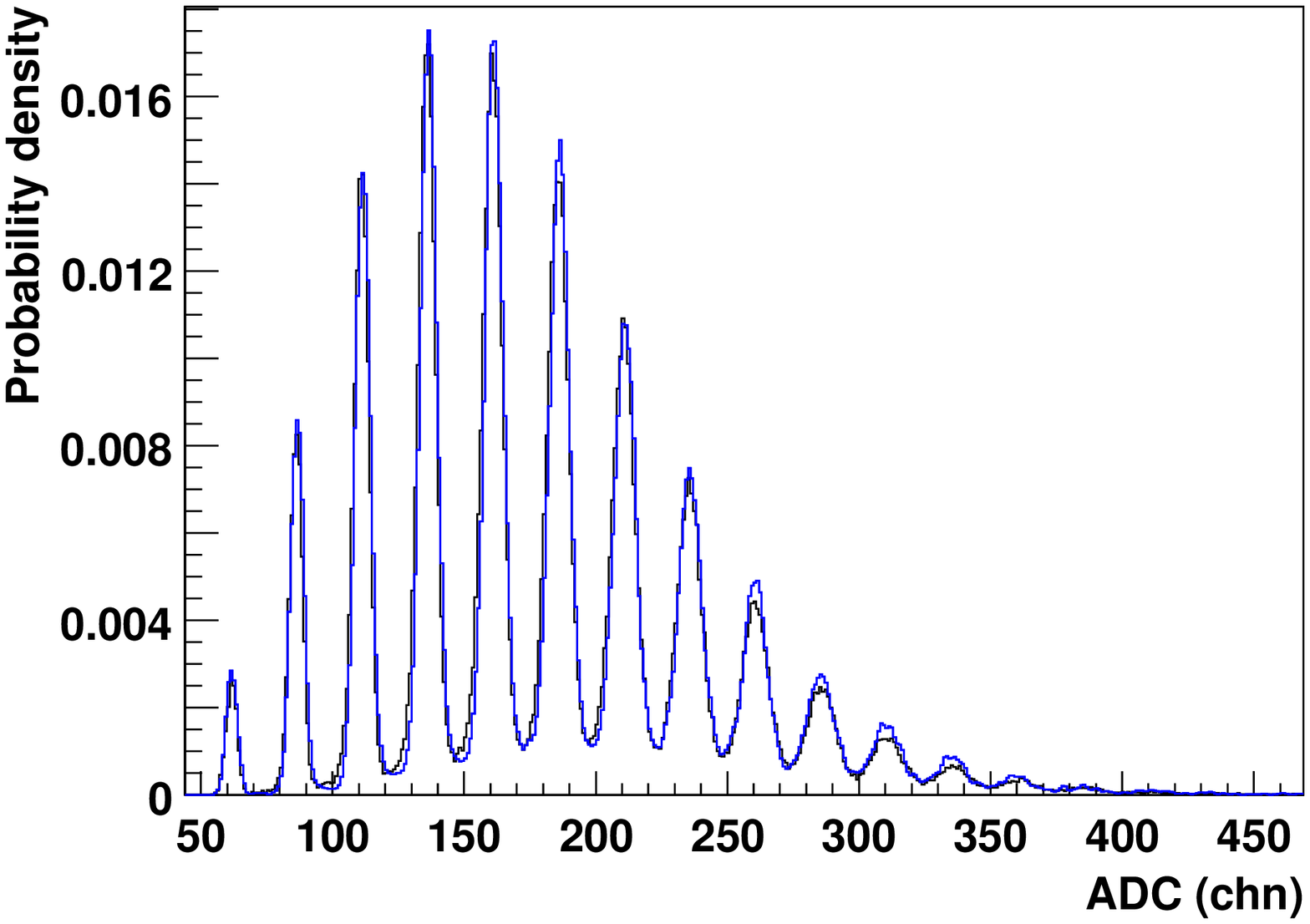}}\hfill
  \subfigure[18.4\,V]{
    \includegraphics[width=0.48\textwidth]{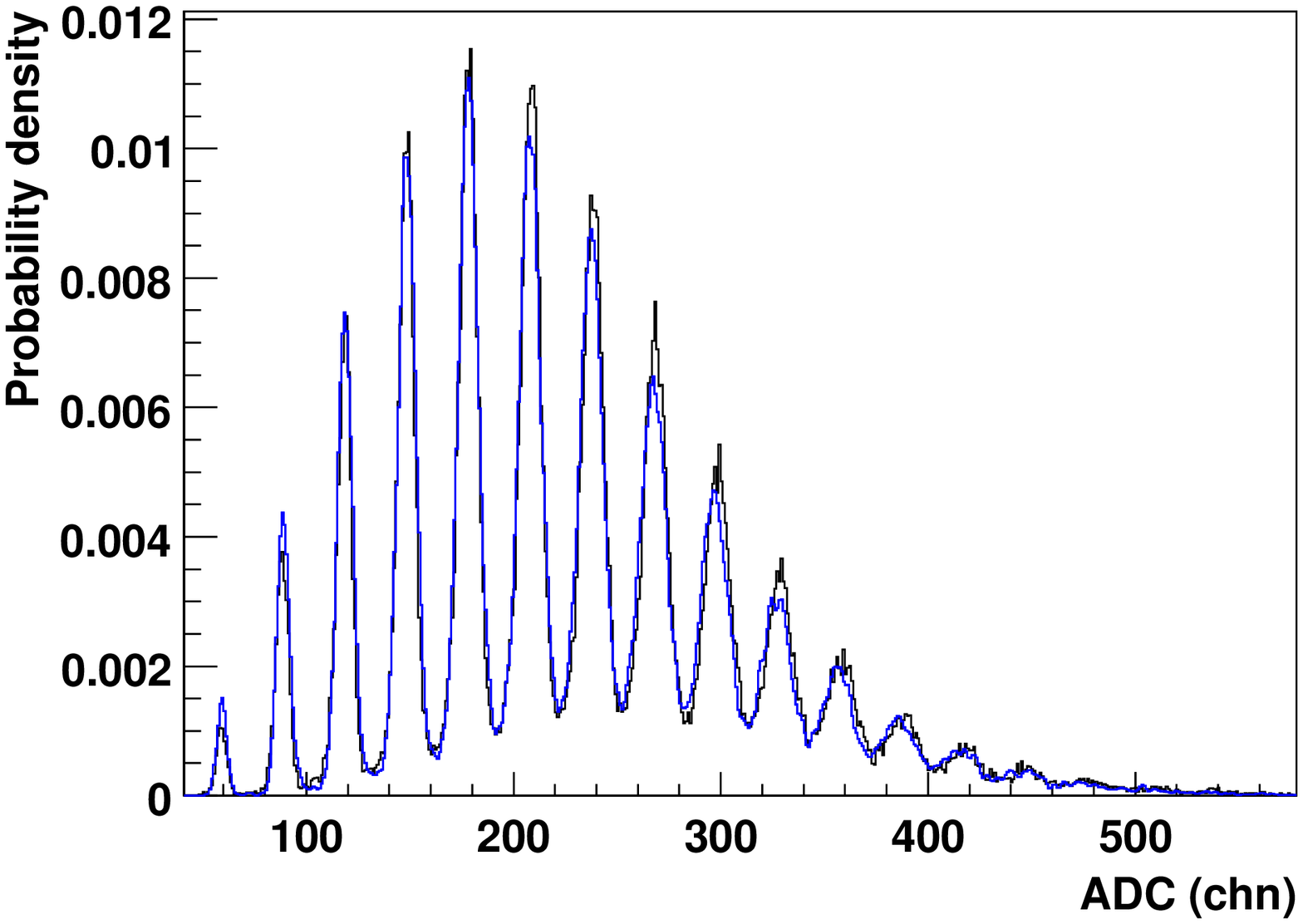}}\\
  \subfigure[18.9\,V]{
    \includegraphics[width=0.48\textwidth]{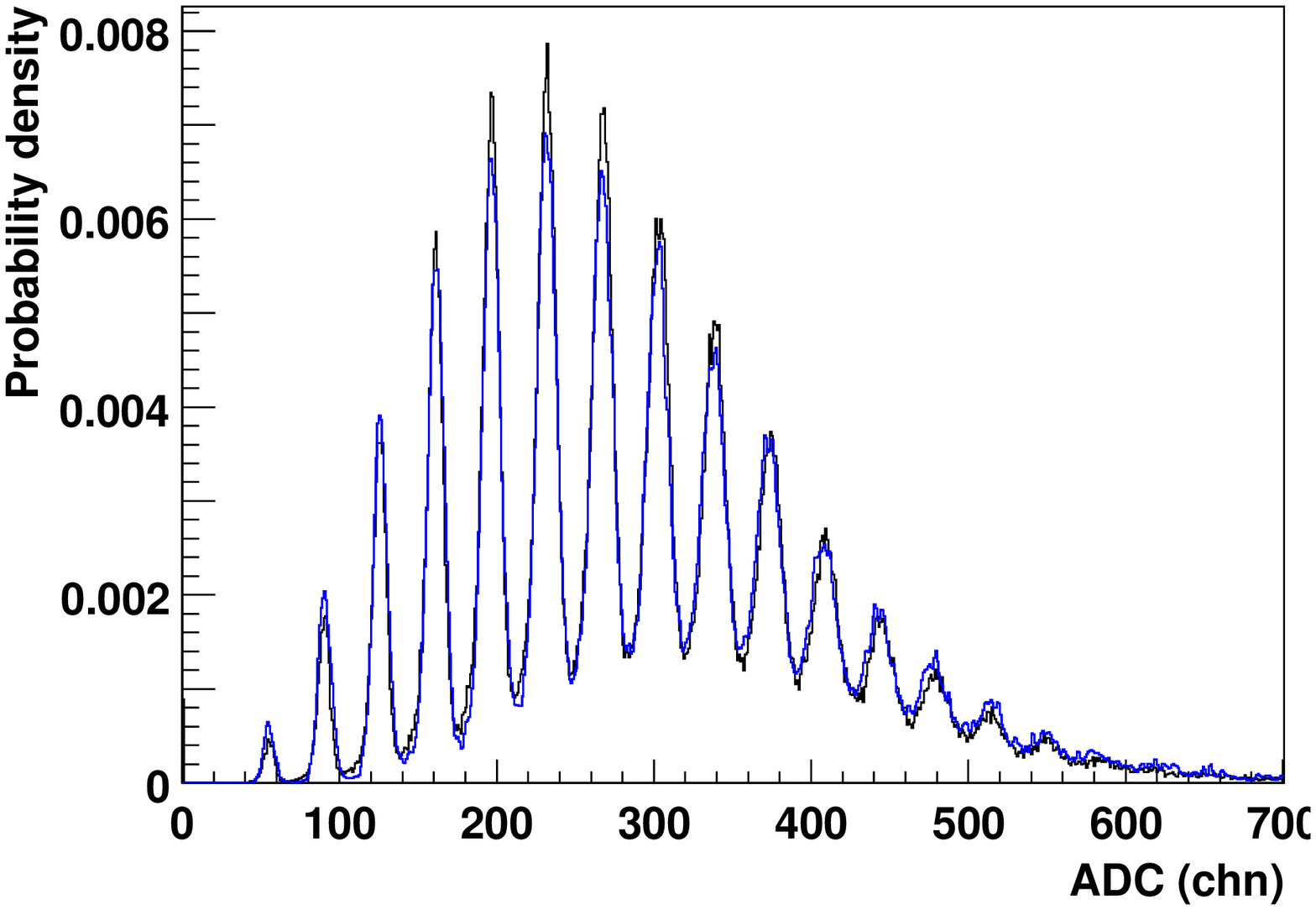}}
  \subfigure[Residuum data-model for all 5 bias voltages]{
    \includegraphics[width=0.48\textwidth]{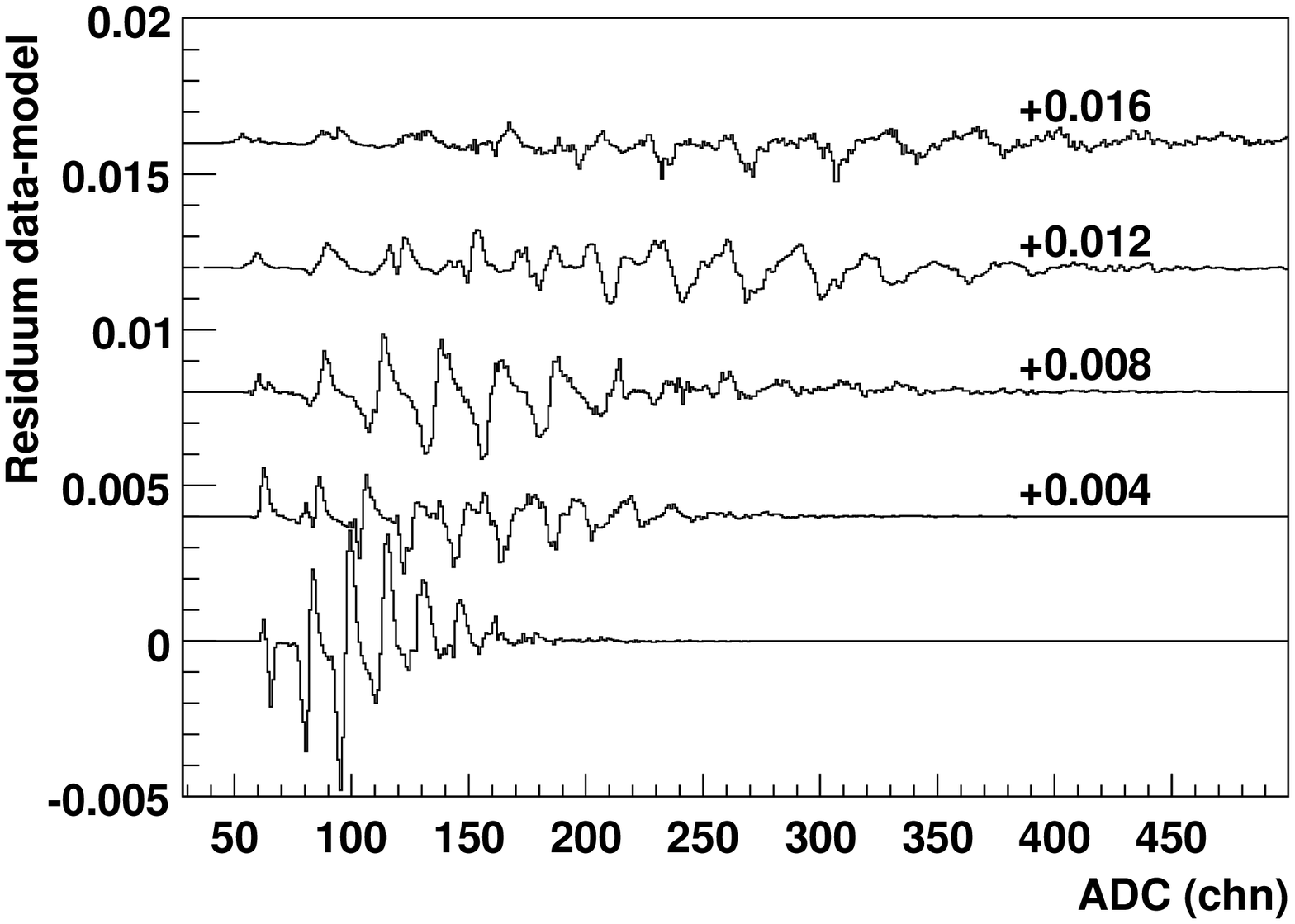}}
  \caption{Measured pulse-height spectra for five different bias
    voltages. Underlying curves are the result of the MPHD model 
    described in the text, see
    Table~\ref{tab:fitpar} for the resulting set of parameters. 
    Panel~(f) shows the residuum between model and data, where
    the curves for difference bias voltages are off-set to improve
    visibility.}
  \label{fig:biasvolts}
\end{figure}
\begin{table}
  \caption{MPHD model results for five bias voltages. The
    free parameters of the simulation are: mean signal amplitude ($A$),
    gain variation ($\sigma_{G}$), dark pulse rate ($r$), optical cross-talk
    probability ($p_{\it opt}$), after-pulse probability ($p_{\it aft}$),
    trap lifetime ($\tau$), mean number of detected photons
    ($\lambda$), pedestal position ($x_{ped}$),
    and noise amplitude ($\sigma_{ped}$).} 
  \centering
  \begin{tabular}{llllll}
  \hline
  \hline
  parameter & 17.0\,V & 17.4\,V & 17.9\,V & 18.4\,V & 18.9\,V \\
  \hline
  $A$ (mV) & 35.&43.&55.&66.&78.\\
  $\sigma_{G}$ (chn)& 0.08 & 0.07&0.05&0.07&0.06\\
  $r$ (MHz)& 3.21& 4.06&4.50&6.35&7.33\\
	$p_{opt}$&0.011 &0.020&0.028&0.033&0.048\\
	$p_{aft}$&0.02 &0.08 & 0.10 &  0.13& 0.17 \\
  $\tau$ (ns)&  8.1 & 8.0 &9.6&8.6&11.9\\
  $\lambda$ (ph.)& 2.76 & 3.27&3.99&4.60&5.06\\
  $x_{ped}$ (chn) & 65.2 & 63.9&61.6&59.0&54.9\\
  $\sigma_{ped}$ (chn)& 1.49 & 1.69&2.10&2.45&3.40\\
  \hline
  \hline
  \end{tabular}
  \label{tab:fitpar}
\end{table}
\begin{figure}
  \centering
  \includegraphics[width=0.8\textwidth]{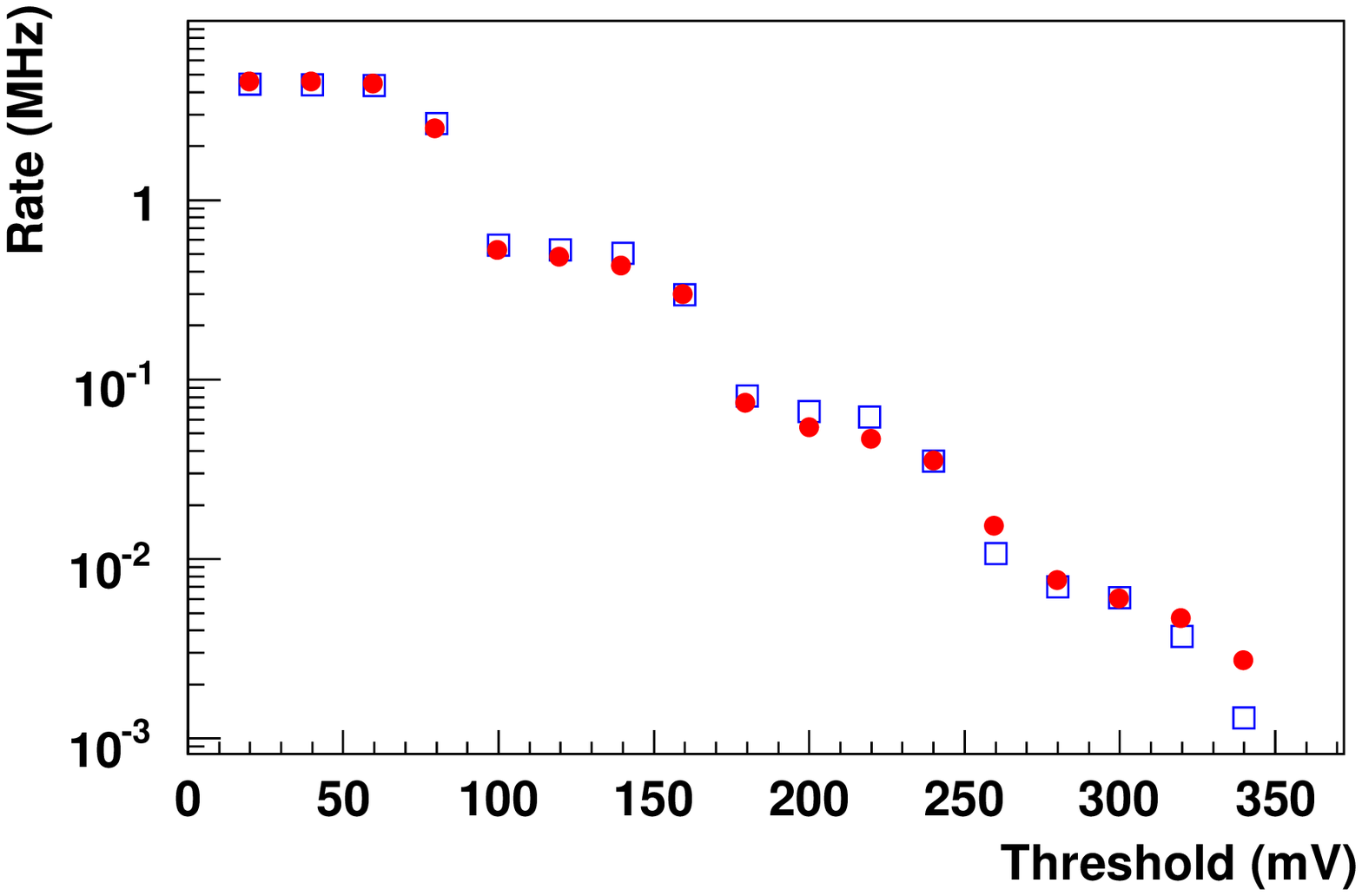}
  \caption{Measured noise rate as a function of discriminator
    threshold for a bias voltage of 17.9\,V. The open symbols are
    the result of the MPHD model. Simulated noise amplitude and gain
    variation have the same values as given in Table~\ref{tab:fitpar}.}
  \label{fig:noiserate}
\end{figure}

Fig.~\ref{fig:biasvolts} shows the measured pulse-height spectra for
low amplitude signals induced by a short pulsed UV
laser\footnote{NanoLED by {\sf Horiba Jobin Yvon},
  http://www.jobinyvon.com/NanoLED (2007)} exciting a plastic
scintillator. The data was taken with five different bias
voltages. Underlying curves are the results of the MPHD model. The
residuum between MPHD model and data is shown in panel (f), where the
curves for difference bias voltages are off-set by $n \times$ 0.004 to
improve visibility. Table~\ref{tab:fitpar} summarises the resulting
set of parameters. Fig.~\ref{fig:noiserate} shows the measured noise
rate for a bias voltage of 17.9\,V. The step structure is due to the
multi-pixel events caused by optical cross-talk. Its sharpness is
governed by the gain uniformity over the pixels and by the narrowness
of the single pixel response function. Last steps are less defined due
to the higher number of pixels involved. The pixels are better
resolved with increasing bias voltages as a result of well known
increase in the charge gain. The strong increase of noise rate with
bias voltage is also visible. The open symbols are a result of the
Monte Carlo. Simulated noise amplitude and gain variation have the
same value as given in Table~\ref{tab:fitpar}.

\section{SiPM irradiation}
A large increase of dark rate has been reported for relatively small
radiation doses~\cite{Musienko}. After irradiation the noise rate
increases because of the introduction of generation centers. This is
the same behavior as for the bulk leakage current in diodes.  In
addition the creation of trapping centers is expected to increase the
after-pulse probability.  The application of the MPHD model described
in the last section for the extraction of the characteristic
parameters is considerably simpler when the multi-photoelectron peak
structure is easily distinguishable. On the other hand the complete
loss of this peak structure defines the limit of the application of
SiPM for low light levels detection. The first irradiation dose was
chosen so that only a small distortion of the peak structure was
obtained. 14\,MeV electrons were used to irradiate a sample of
SSPM-0701BG-TO18 diodes. The beam current was 10\,nA. The electrons
crossed a 0.3\,mm thick aluminum window at 15\,cm distance from the
1\,mm$^2$ active area of the SiPM. Fluences on the diodes ranged from
$3.1 \times 10^9$ to $3.8 \times 10^{10}$\,electrons$/$mm$^{2}$. Heat
dissipation and damage on the transparent epoxy layer protecting the
silicon material were calculated and proved to be
negligible. Grounding of the diode was provided in order to avoid
damage by the sudden release of accumulated charge.

\section{Characterisation of the irradiation damage}
\begin{figure}
  \centering
  \includegraphics[width=0.8\textwidth]{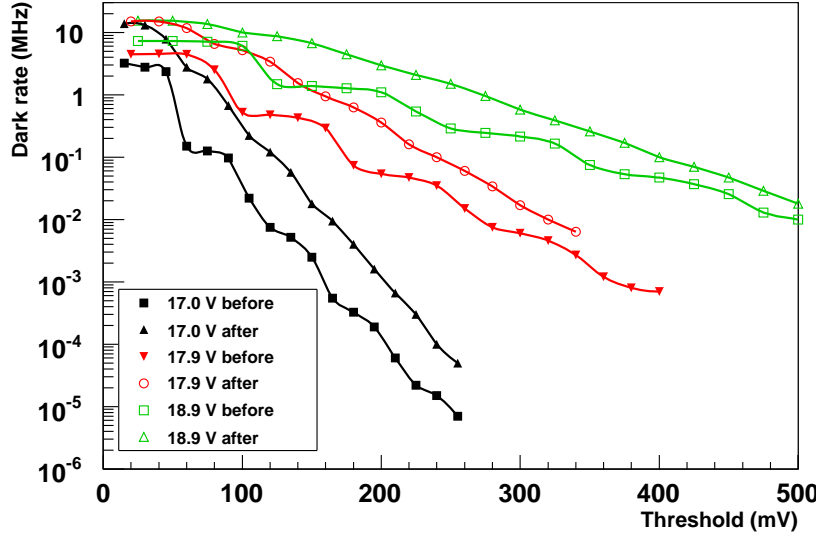}
  \caption{Measured noise rate before and after irradiation with
    3.1$\times 10^8$\,electrons$/$mm$^2$ of 14\,MeV energy as a
    function of threshold in a leading edge discriminator. The plot
    shows curves for three different bias voltages. The step structure
    is due to the multi-pixel events caused by optical
    cross-talk. After irradiation the steps are less pronounced
    because of the much higher noise rate.}
  \label{fig:noiserate-irrad}
\end{figure}

The SiPM were characterised before and after irradiation by studying
the noise rates and the pulse-height spectra for low amplitude
signals. Fig.~\ref{fig:noiserate-irrad} shows the noise rate before
and after irradiation with $31 \times 10^8$ electrons of 14\,MeV
energy as a function of threshold in a leading edge discriminator. The
figure includes curves for three different bias voltages.  Two
observations were made after the irradiation:
\begin{enumerate}
  \item the rates of dark pulses are significantly larger.
  \item the steps are much less pronounced than before irradiation.
\end{enumerate}
The simulation shows that an increase in the dark count
rate is insufficient to explain the curves and it confirms that either
an increase in the noise amplitude or the loss of gain uniformity (or
a combination of both) can reproduce the measured values.

\begin{figure}
  \centering
  \subfigure[bias voltage of 17.0\,V]{
    \includegraphics[width=0.48\textwidth]{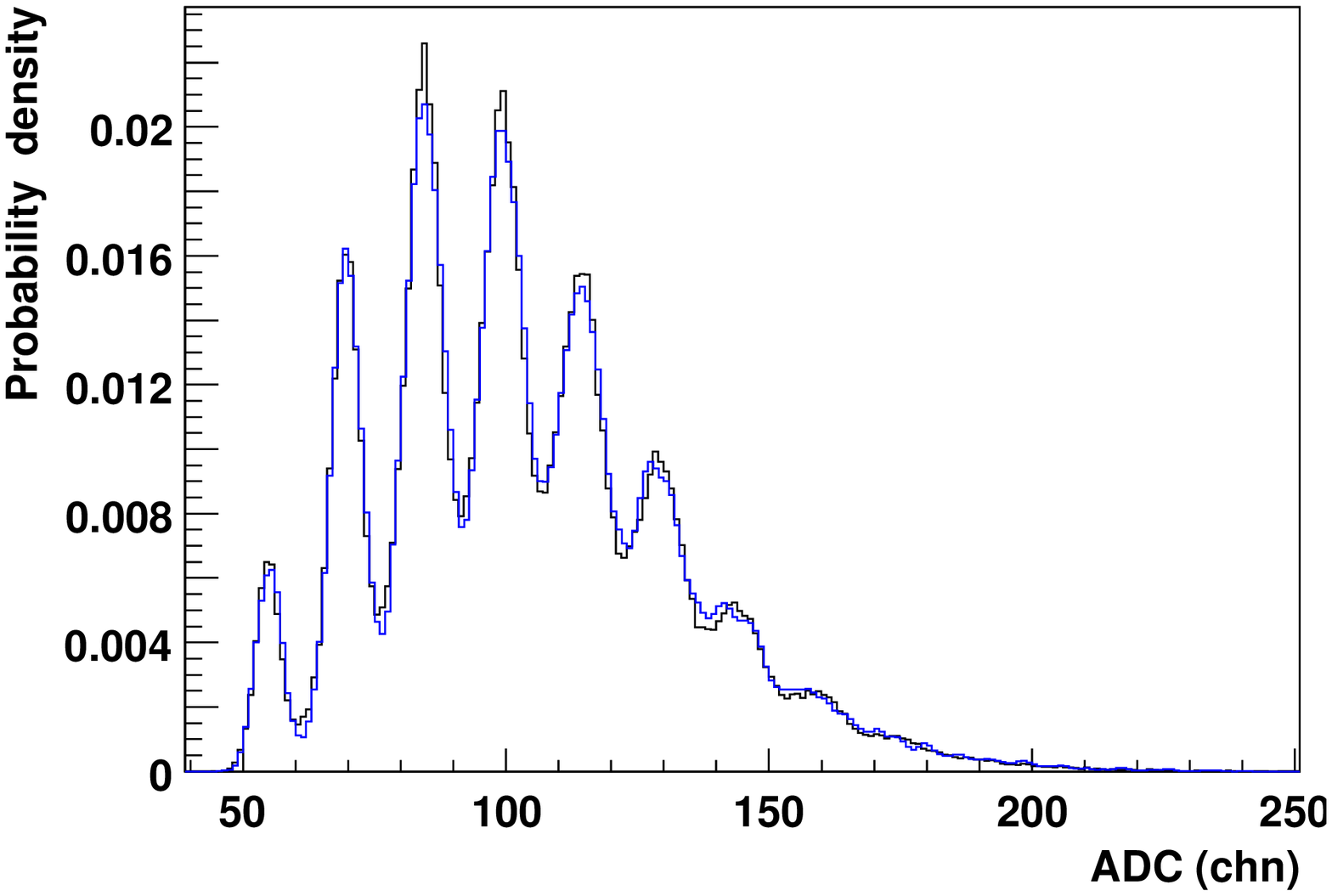}}
  \subfigure[bias voltage of 17.4\,V]{
    \includegraphics[width=0.48\textwidth]{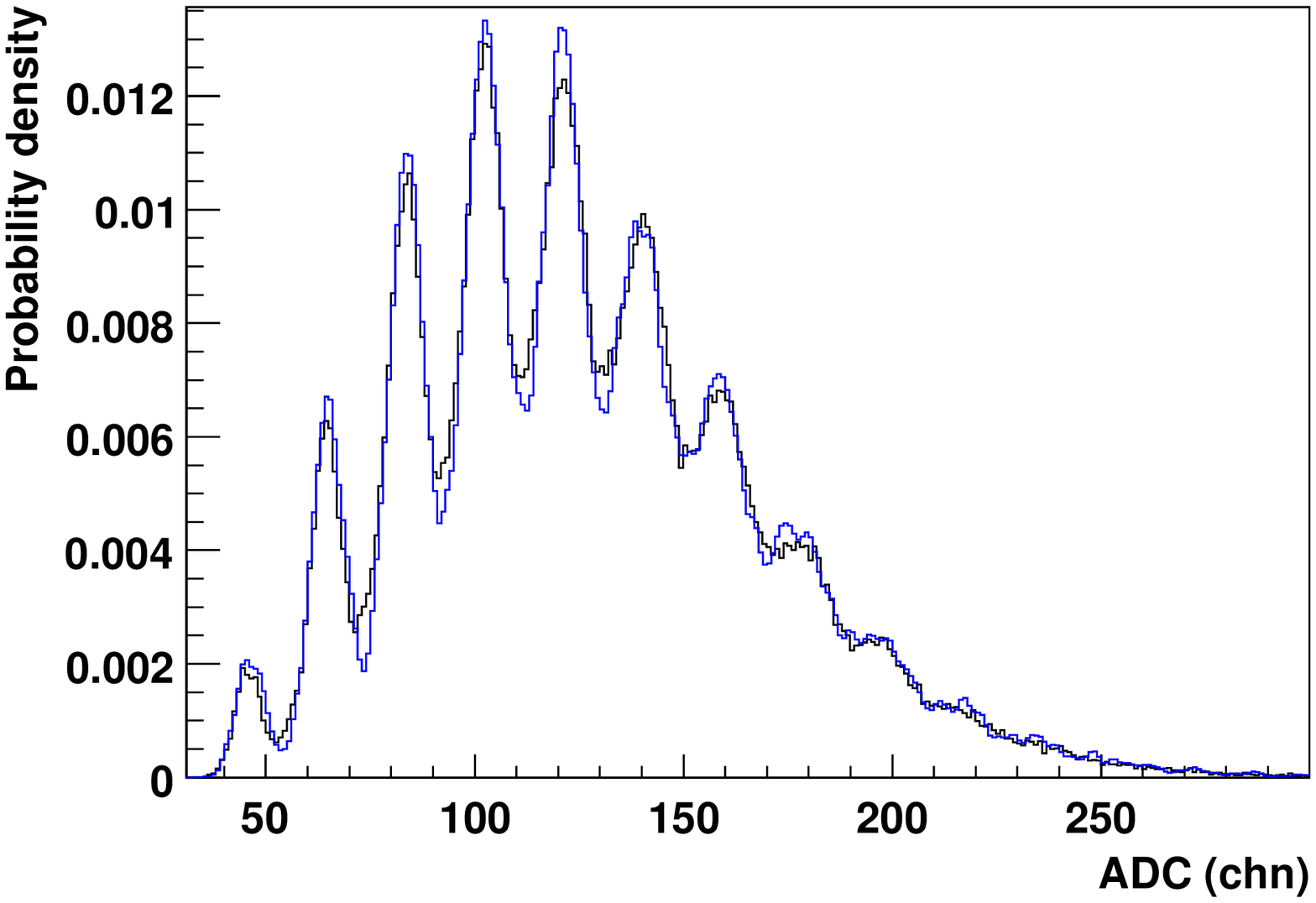}}
  \caption{Pulse-height spectra after irradiation with 3.1$\times
    10^8$\,electrons$/$mm$^2$ of 14\,MeV energy for low amplitude
    signals. The underlying curves are a result of the MPHD model.}
  \label{fig:radiated}
\end{figure}

The histograms in Fig.~\ref{fig:radiated} show measured ADC spectra
for bias voltages of 17.0 and 17.4\,V (compared to 17.9\,V recommended
voltage).  The MPHD model was tested on this low photon yield data.
The simulated integration of signals was performed in an interval of
30\,ns, and the mean width of the single-pixel signal was 9\,ns.  The
pedestal peak in this case was clearly separated from the one-pixel
peak. After-pulse probability was set to zero and the values for the
noise rate and optical cross-talk probability were deduced from the
measured noise rates as a function of discriminator threshold. The
value of $q$ was given to a good approximation by the ratio of noise
rates above and below the first step $P(2)/P(1)= 4(q^3-2q^2+q)$.  The
dark count rate might naively be taken from the measurement at lowest
discriminator threshold, however, this value is not reliable due to
the high probability of signal pile-up.  If one assumes that the
optical cross-talk probablility was not significantly modified by the
low radiation doses the dark count rate can be calculated by
multiplying the noise rate for a threshold between one and two pixels
by $1/4(q^3-2q^2+q)$.  The distributions were well reproduced when
noise amplitude and gain variation were allowed to change. The values
for the gain variation before and after irradiation were 0.08\,chn and
0.12\,chn, respectively.  The noise amplitude changed from 1.49\,chn
before to 2.44\,chn after irradiation. The good matching of the two
curves is a confirmation of the hypothesis of low after-pulse
probability.  It seems necessary to conclude that there has been an
increase in leakage current and a severe loss of gain uniformity.
  
\begin{figure}
  \centering
  \subfigure[Pulse-height spectrum]{
    \includegraphics[width=0.48\textwidth]{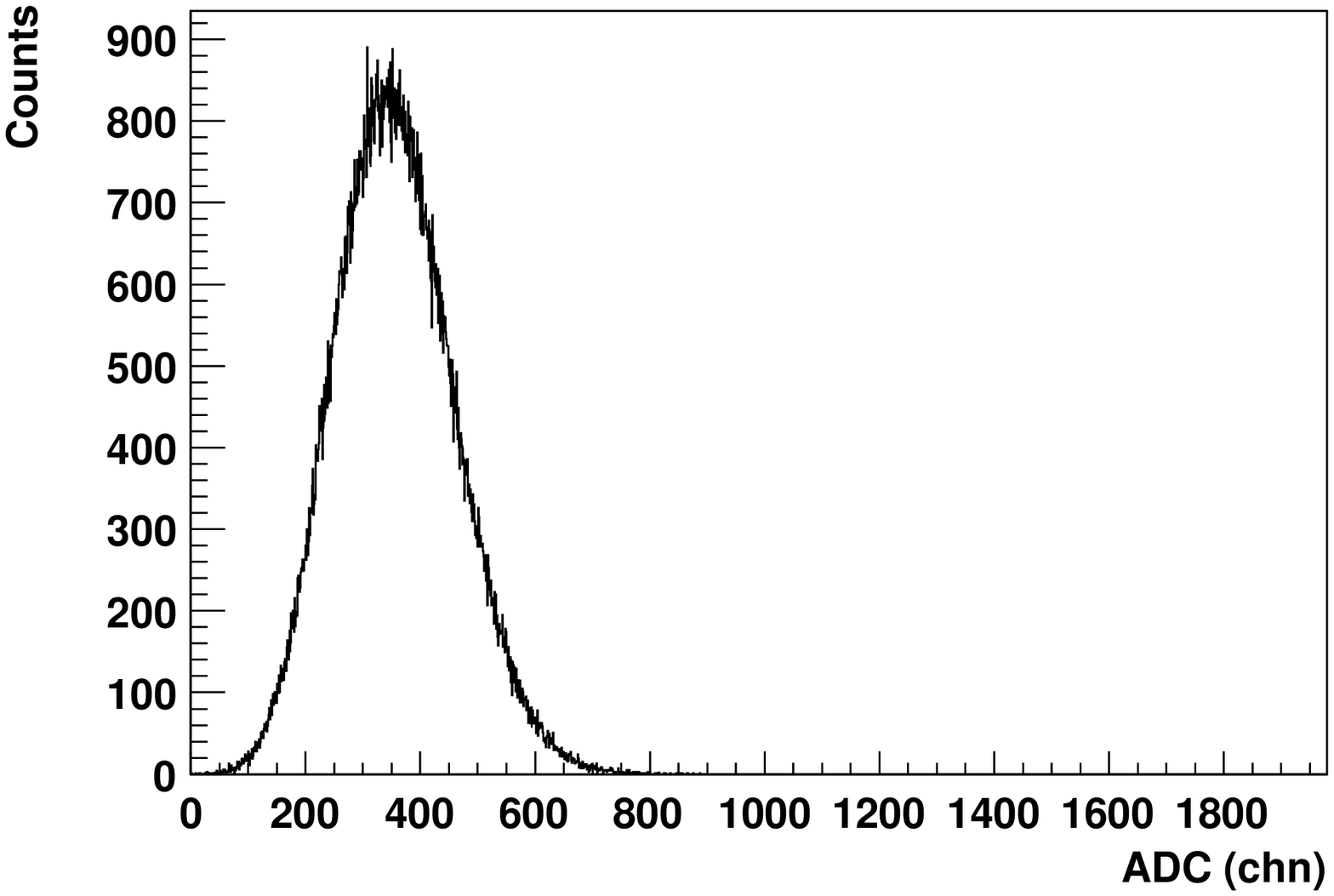}}
  \subfigure[Pulse-height peak position]{
    \includegraphics[width=0.48\textwidth]{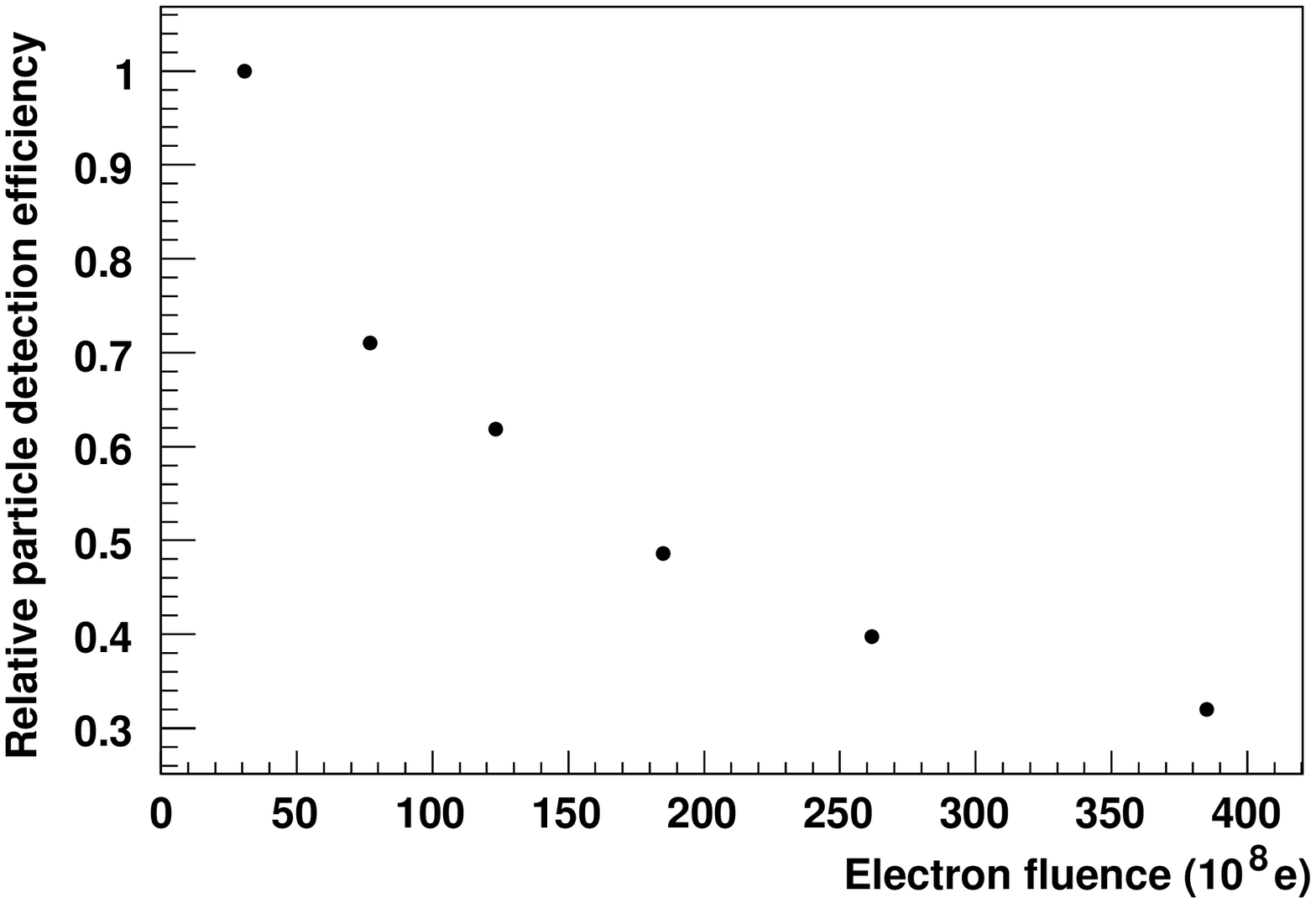}}
  \caption{(a) Pulse-height spectrum obtained with larger laser
    intensity. Individual peaks are no longer visible. The ADC channel
    corresponding to the maximum probability is a good measure of
    photon detection efficiency. (b) Relative photon detection
    efficiency for increasing electron fluence.}
  \label{fig:highrad}
\end{figure}

For higher fluences the noise increase is so large that a
multi-photoelectron peak differentiation is no longer possible. A
different approach was used for dealing with such a situation: noise
should not be relevant for large signals but a variation in the gain
or photon detection efficiency should still be noticeable by studying
the position of the maximum in the pulse-height spectrum, see
Fig.~\ref{fig:highrad}\,(a).  The result of such an investigation is
shown in Fig.~\ref{fig:highrad}\,(b). The single pixel signal was
monitored after each irradiation and no change in its amplitude was
observable. The conclusion has to be that there has been a progressive
reduction of the photon detection efficiency.  This fact can be
attributed to the loss of a progressively larger amount of pixels
which remain permanently in the off state or to the reduction in the
photon detection efficiency of each pixel.  Any of these two effects
would be equally problematic for low light level detection.

\section{Conclusions}
Multi-anode photomultipliers are routinely used for scintillating
fibre detector read-out showing all the drawbacks of conventional
phototubes: stiff cables, high voltage power supplies and magnetic
field sensitivity in addition to the specific problem of optical
cross-talk among their many channels. If SiPM could be used instead an
overall price reduction and a considerable detector simplification
would be obtained.  Conventional experimental techniques used to deal
with high noise rates such as coincidences of several detectors can be
complemented with more sophisticated methods based on intelligent
trigger algorithms implemented in FPGA chips in order to obtain a
reliable tracking detector. Cooling Peltier modules would reduce
dramatically the noise rates in case the mentioned methods prove to be
insufficient~\cite{Eraerds}.

For the \KAOS\ spectrometer at the Mainz Microtron, Germany, the
viability of a fiber detector is studied as part of the electron arm
tracking system in which SiPM are considered as possible photon
detectors~\cite{Achenbach-SNIC06}.  Simultaneously, a scintillating
fibre tracker with SiPM read-out is being considered for the
time-of-flight start detector or for a secondary active target in the
future \PANDA\ experiment~\cite{PANDA} at {\sf FAIR}.  Here, radiation
hardness is an important issue due to the relatively small distance
from the target area to the detector position.

A fluence of only $1.7 \times 10^{10}$ particles reduces already the
number of detected photons by a factor of two.  Good shielding will be
necessary in many applications where these doses are accumulated in a
relatively short time.
  
\section*{Acknowledgements}
We thank the MAMI accelerator staff for their help in the 
SiPM irradiations. Work supported in part by the European
Community under the ``Structuring the European Research Area''
Specific Programme as Design Study DIRAC\-secondary-Beams (contract
number 515873).

%%%%%%%%%%%%%%%%%%%%%%%%%%%%%%%%%%%%%%%%%%%%%%%%%%%%%%%%%%%%%%%%%%%%%
%                         BIBLIOGRAPHY                              %
%%%%%%%%%%%%%%%%%%%%%%%%%%%%%%%%%%%%%%%%%%%%%%%%%%%%%%%%%%%%%%%%%%%%%

%
%\bibliographystyle{elsart-num}
%\bibliography{sipm}

\end{document}